\documentclass[10pt]{article}

\usepackage[utf8]{inputenc}

\usepackage{listings} % for showing python scripts
\usepackage{epsfig}
\usepackage{graphicx}
\usepackage{amsmath}
\usepackage{amsfonts}
\usepackage{rotating}
\usepackage{authblk}
\usepackage{float}
\usepackage{lscape}
\usepackage[english]{babel}
\usepackage{setspace}
\usepackage{amssymb}
\usepackage[a4paper, margin=1in]{geometry}
\usepackage{color}

\newcommand{\be}{\begin{eqnarray}}
\newcommand{\ee}{\end{eqnarray}}
\newcommand{\ba}{\begin{array}}
\newcommand{\ea}{\end{array}}
\newcommand{\bi}{\begin{itemize}}
\newcommand{\ei}{\end{itemize}}
\newcommand{\bv}{\begin{verbatim}}
\newcommand{\ev}{\end{verbatim}}
\newcommand{\bt}{\begin{tabular}}
\newcommand{\et}{\end{tabular}}
\newcommand{\btab}{\begin{table}}
\newcommand{\etab}{\end{table}}
\newcommand{\bfig}{\begin{figure}}
\newcommand{\efig}{\end{figure}}
\newcommand{\bc}{\begin{center}}
\newcommand{\ec}{\end{center}}
\newcommand{\bit}{\begin{itemize}}
\newcommand{\eit}{\end{itemize}}
\newcommand{\nn}{\nonumber}

\newcommand{\Sum}{\displaystyle\sum\limits}
\newcommand{\tw}{\textwidth}
\newcommand{\ig}[1]{\includegraphics[width=#1\tw]}

\newcommand{\bmnp}{\begin{minipage}}
\newcommand{\emnp}{\end{minipage}}

\title{Evidence of Fraud in Brazil's Electoral Campaigns Via the Benford's Law}
\author[a]{Daniel Gamermann}
\author[a]{Felipe Leite Antunes}
\affil[a]{Universidade Federal do Rio Grande do Sul (UFRGS) - Instituto de Física, Av. Bento Gonçalves 9500, Porto Alegre, Rio Grande do Sul}

\begin{document}
\maketitle

\abstract{The principle of democracy is that the people govern through elected representatives. Therefore, a democracy is healthy as long as the elected politicians do represent the people. We have analyzed data from the Brazilian electoral court ({\it Tribunal Superior Eleitoral, TSE}) concerning money donations for the electoral campaigns and the election results. Our work points to two disturbing conclusions: money is a determining factor on whether a candidate is elected or not (as opposed to representativeness); secondly, the use of Benford's Law to analyze the declared donations received by the parties and electoral campaigns shows evidence of fraud in the declarations. A better term to define Brazil's government system is what we define as chrimatocracy (govern by money).}

\emph{keywords: }{Benford's Law, Logistic regression, Electoral campaign, Politics, Fraud}

%\justifying

%%%%%%%%%%%%%%%%%%

\section{Introduction}\label{introdution}

Modern society dependence on technologies, in particular the Internet and mobile phones, has as consequence the generation of huge amounts of raw data. Apart from the problematic involved in the processing and storage of this data, the data's volume, structure and variety call for the development of new analysis methodologies in order to extract the important information (knowledge) behind it. Also, as scientific fields that have traditionally adopted qualitative approaches slowly tackle quantitative analyses, a vast new horizon opens to new applications of methodologies long known to the physics community. 

This interaction of physics with other sciences has been fruitful in apparently distant fields such as economics \cite{econ, Stanley2002, Amaral1999, Stanley1995}, biology \cite{bio}, medicine \cite{onias2013} or political sciences\cite{Filho1999, extreme}. In this context, Statistical Physics has much to offer, particularly in understanding, quantifying and modeling the dynamics and properties of a large number of elements. Big Data \cite{bigData} with its unprecedented scale and much finer resolution, provides a powerful experimental apparatus to challenge our existing models, explore new tools and frameworks, and lead research to new areas \cite{wang}.

At the moment, the sector that most benefits from the rising data science field is the private sector. Companies invest heavily in studding costumer profiles and needs in order to offer more attractive services and increase their profits or optimizing decision making process minimizing risks. On the other hand, the public sector might enormously benefit from knowledge obtained with these new information technologies. Objective analysis could guide public policies preventing the spread of epidemics \cite{epidemia,Gonzalez2004}, minimize traffic jams\cite{Sauermann1998}, decreasing unemployment \cite{desemprego}, fighting corruption \cite{corrupt1, corrupt2, Podobnik2015}, crime \cite{crime2, crime} or violence \cite{violence}.

An interesting result applied in the detection of fraud is the Benford's Law. Noted for the first time by the astronomer and mathematician Simon Newcomb \cite{newcomb}, and empirically postulated by Benford when comparing data collected from a variety of sources, ranging from the statistics of the American baseball league to the atomic weights of the elements, the law of probability of occurrence of numbers, as observed by Newcomb, is such that all the mantissas \footnote{The mantissa of a given number $ x $ is such that $ x = m \times 10 ^ n $, Where $ n $ is an integer.} of its logarithms are equiprobable. This observation can be put as follows \footnote{More accurately, the uniform distribution of the log of the mantissa is equivalent to the generalized Benford's distribution for n-digits \cite{Shao2010}.} \cite{benford}:

\be
P(d) &=& \log_{10}(d+1)-\log_{10}(d) = \log_{10}\left(1+\frac{1}{d}\right), \\
d &\in& \{1, 2, .., 9\}.
\ee
Despite its simplicity, the first rigorous proof was only developed by Hill in 1995 \cite{hill1995}. In the original work, Hill proves, based on probability theory, that scale invariance implies base invariance and base invariance, in turn, implies the Benford's Law.

Sets of numbers tend to follow this law given that they are naturally occurring (random) numbers, coming from multiple different distribution and expanding many orders of magnitude. By naturally occurring numbers, it is meant numbers that are not sequential, man made, as would be for example, serial numbers or license car plates, which would not be random, but cover a given range uniformly. It is interesting to note that this law is scale invariant, so it does fell as a natural law (independent of man made measurement systems or concepts): i.e. take the measurement of the heights of all mountains in a country, if they tend to follow Benford's law, they will do so no matter if the measurements are made in meters, feet or inches. The distribution of the first digit will have approximately the same shape no matter the unit system used. Were the distribution uniform in a given measurement system, it would have a complete different shape in another system, the distribution would then be measurement system dependent.

Benford's law may be an important tool in order to search big amounts of data for anomalies. It is interesting to note that Benford's law has already been used in order to detect evidence of fraud in electoral results \cite{iran} and in revenue tax declarations \cite{revenue, durtschi}. Being an important accounting forensic tool it could be admissible as evidence in courts of law.

In this work we analyze publicly available data on Brazilian elections. Brazil's superior electoral court (TSE from {\it Tribunal Superior Eleitoral}) freely provides all statistics on election results and financial declarations made by parties, candidates and electoral committees. This information can be downloaded from the TSE webpage \cite{TSE} (see also the appendix).

Ideally, in a democracy, the people elects its leaders based on representativeness. Those politicians that better represent the population or groups within the population and better defend their interests should end up elected. The electoral campaign is the opportunity the candidates to offices have to express their ideas and the voters to get acquainted with the candidates and to chose those that better represent their interests. In practice, Brazil's system faces many problems. On one hand, not all candidates have the same opportunity to appear in front of the population and express their plans; on the other hand, no matter what a politician promises during the campaign, once elected he can follow a completely different line. The first problem, we believe, can be traced to a single factor: money. Electoral campaigns are much closer to plain publicity than to ideological debate. The more money a candidate or a party has, the better the marketing professionals he can hire and the more time he can buy in private media and consequently, the more he is remembered by the voters. The public media time is shared by the candidates and parties, but it is proportional to the number of congressmen each party has, such that one has a positive feedback effect: the more time a party has, the bigger the opportunity it has to influence the voters, therefore, the bigger the probability it has to elect its members and the bigger the media time it will have in the next election. It is easy to realize the nasty effect money has in an electoral campaign, completely perverting the principles of democracy. At this point, we would like to define the term Chrimatocracy\footnote{The fact that this word appears to have its roots in the word crime is just a happy coincidence.}, from the greek word $\chi\rho\eta\mu\alpha\tau\alpha$ which means money. Chrimatocracy is the system of government where the ones who receive more money, govern. The principle of democracy would therefore be broken: in a country where the majority of the population is relatively poor, those who have big amounts of money to donate to politicians do not represent the people.

After obtaining some descriptive statistics on the data from Brazil's TSE, two analyzes are performed: using logistic regression we determine the relationship between the money a candidate declares he received as donations and the probability of him to get elected for office; in a second analysis, we study the set of all single donations received by each player (party, candidate or committee) and search it for anomalies not following Benford's law. We evaluate the statistical significance of this discrepancy and we also construct a random model for donations and create random sets of donations with similar descriptive statistic than the declared donations to perform the same test over the modeled sets of numbers. In the next section we describe the data and the analyses performed, in the section after that we present and discuss our results and in the last section we give a short overview and present our conclusions.

%%%%%%%%%%%%%%%%%

\section{Materials and Methods}

In this section, the data used is explained. All data used in this study is publicly available from Brazil's superior electoral court (TSE). In the appendix, we describe how to obtain the data from Brazil's TSE, exactly which files were used and how to download them. Based on the data statistics, a model is elaborated in order to generate artificial data to compare the results when performing the Benford analysis. Last in this section we briefly explain the logistic regression model.

\subsection{Data}

Brazil has elections every two years, but alternating between two different types of elections, each type occurring every four years. There are the municipal elections, where mayors and city council members are elected (the last one occurred in 2016) and general elections where president, governors, senators and congressmen (regional and national) are elected (the last one occurred in 2014). Brazil has 26 federal units plus the federal district. Each one of these units (regions) elects its senators, congressmen and governors.

For each federal unit, Brazil's TSE provides information on the donations declared by the three entities: candidates, parties and committees. The data comprises information describing every donation received. The donations can be divided in two categories with respect to the donor: they can come from legal persons (private citizens, identified by the CPF\footnote{CPF is an identification number used by the Brazilian tax revenue office. It is roughly the Brazilian analogue to a social security number. With the same purpose, companies are identified with a similar number called CNPJ.} number) or from legal entities (i.e. companies, identified by the CNPJ number). Also, some entities can make donations among them (the party can give part of the money from a given donation to a candidate). In this type of transaction, the information on the original donor is also specified in the declarations. From now on, these type of donations will be referred to as non-original donations. Apart from information concerning each Brazilian federal unit separately, one can also obtain the information declared by the parties and committees at national level and for the presidential campaign (which has national and not regional scope).

Parallel to financial information on the donations declared for the electoral campaigns, we also obtained information concerning the elections results (valid votes obtained by each candidate, and his situation: elected or not) and the party coalitions in each federal unit. This information is interesting because for some offices, not necessarily the most voted candidates are elected, but the number of congressmen elected for a given party coalition depends not only on the votes obtained by a single candidate, but all votes to candidates in the coalition determine how many seats the coalition receives and then these seats are distributed among the most voted candidates within the coalition. So, a candidate can have more votes than a minimum needed for being elected and the excess votes somehow go to less voted candidates of his coalition. In practice, some times a candidate is elected having received less votes than some of his non-elected competitors (yes, this is our ``democracy'').

We will present analysis done with data for the 2014 elections. In this election, Brazil's president was elected, along with the national congress, senate and regional governments. When analyzing the data, we do not mix information for national with regional elections. Therefore, first we present three different sets of data: the donations specific for the presidential campaign (donations received by comittees and candidates), donations received directly by the parties (which end up distributed among candidates or committees as non-original donations) and the donations received by the governor regional campaign in one federal unit, the state of Rio Grande do Sul.

In our analyses, the donations can be divided in four categories according to their nature: CNPJ, CPF, Non-original and Unknown (donations for which neither a CPF nor a CNPJ has been attributed). For each set of donations, the distribution of the first digits in the amounts donated are obtained and compared to the Benford's Law by performing a standard $\chi^2$ test, the p-value obtained from this test is the probability that a fluctuation as big or bigger than the observed one comes from a distribution with the assumed shape. So the bigger the p-value, the more the observed frequencies are in agreement with the expectation of Benford's Law.

But then, it is fair to ask, why should the amounts declared as donations have the first digit distribution following Benford's law? Although in some cases a satisfactory explanation for the manifestation of Benford's law in some sets of naturally occurring numbers has eluded mathematicians, explanations have been given for describing this phenomenon in sets of numbers that come from multiples distributions expanding many orders of magnitude \cite{hill1995}. We argue that this is the case with electoral donations. Donations are not made by fixed amounts, they are in principle, spontaneous, the donor chooses the amount he wants to donate. The amounts donated are, in this sense, random and not sequential or uniform. Electoral campaign rules only determine that the maximum amount a legal person can donate should not exceed 10\% of his revenue in the year before election and donations from legal entities should not exceed 2\% of its brute revenue\footnote{Actually, this rule was valid only until 2014 election. After this year, legal entities are (officially) forbidden to donate.}. This characteristic of the donations results in values expanding many orders of magnitude: richer persons or companies can donate more, much more, than poor citizens or small business. In fact, in figure \ref{fig:allamounts} we show the cumulative distribution for all declared donations at national level (not regional), with the horizontal axis in logarithmic scale. This figure clearly shows that a range encompassing 7 orders of magnitude (the smallest donation is 1 real and the maximum is 14000000 R\$) is covered and in the detail, one can see that legal entities donation values are on average much bigger than common citizens donations (multiple different distributions). The statistics for the four distributions is shown in table \ref{tab1} (note, from the table that 11 donations have not been attributed to neither a legal person nor entity). In table \ref{tab:tabcp} we show the descriptive statistics for the donations made to the central directories of the parties. This is the money that the parties have to redistribute among all their candidates and campaigns.

\bfig
\bc
\bt{cc}
\ig{.4}{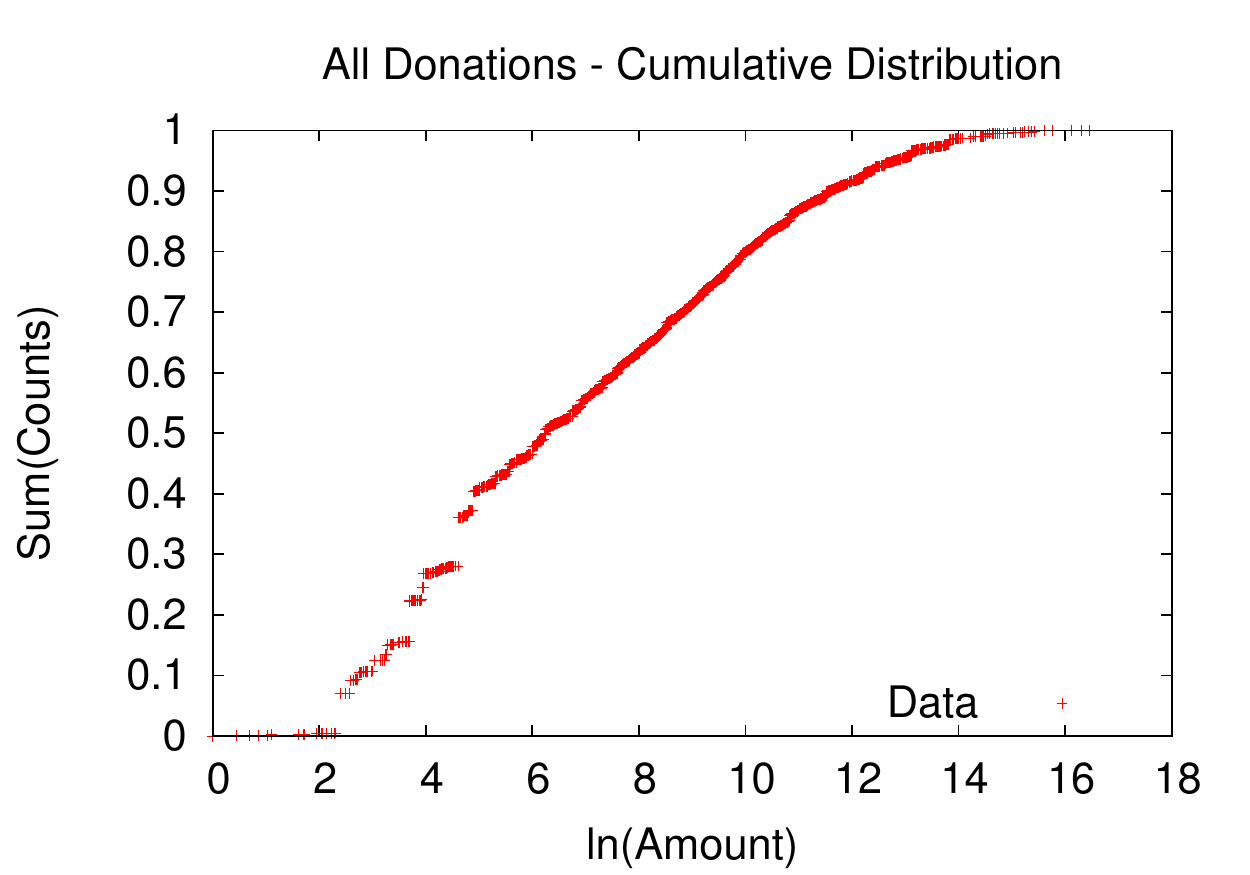} & \ig{.4}{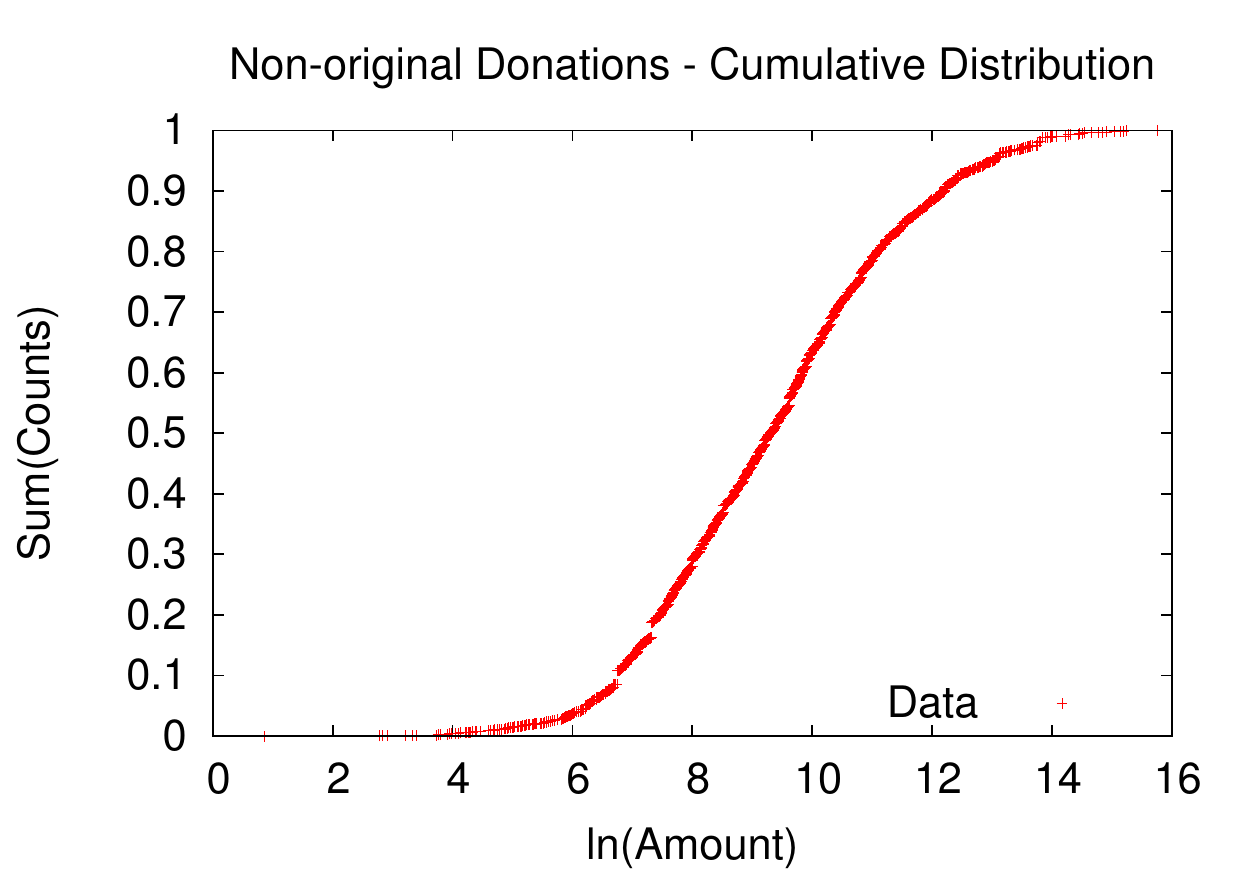} \\
\ig{.4}{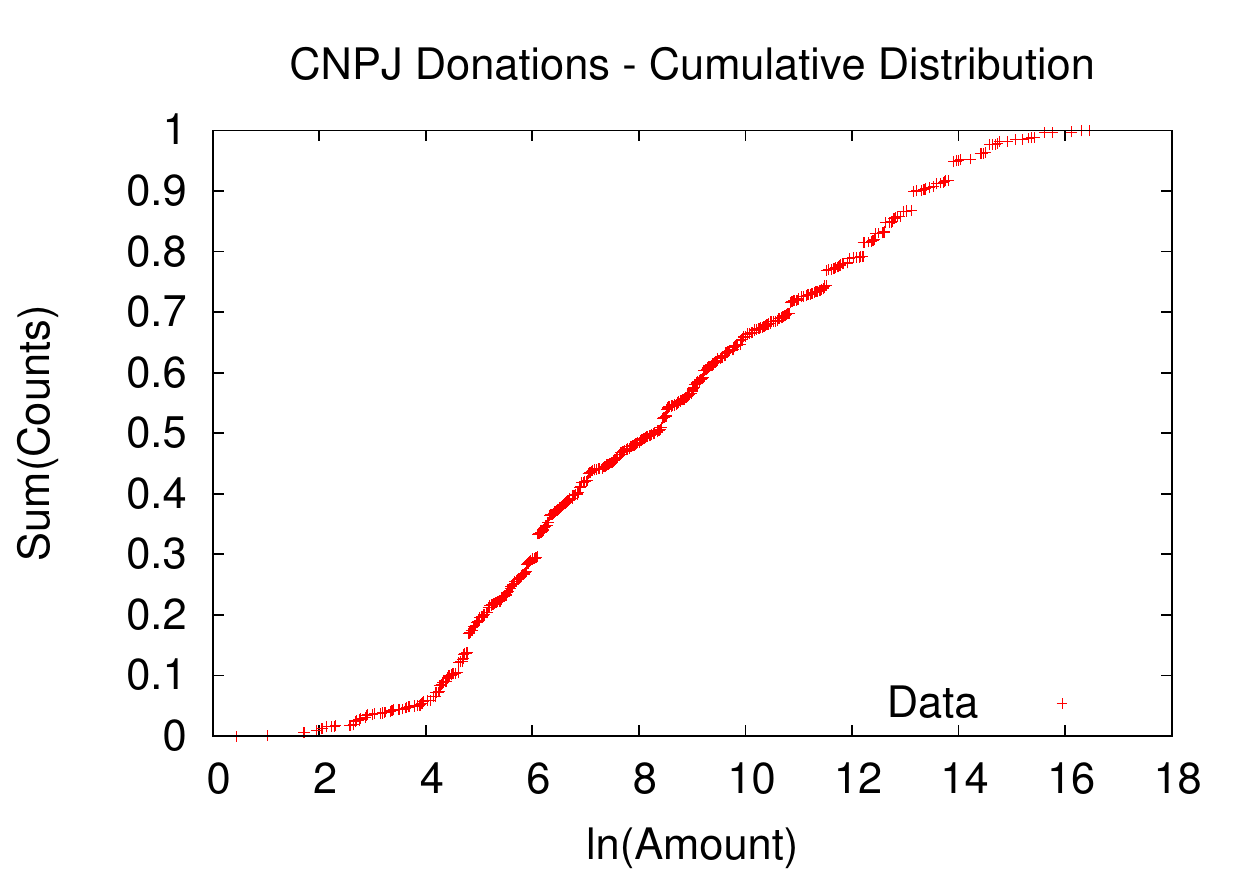} & \ig{.4}{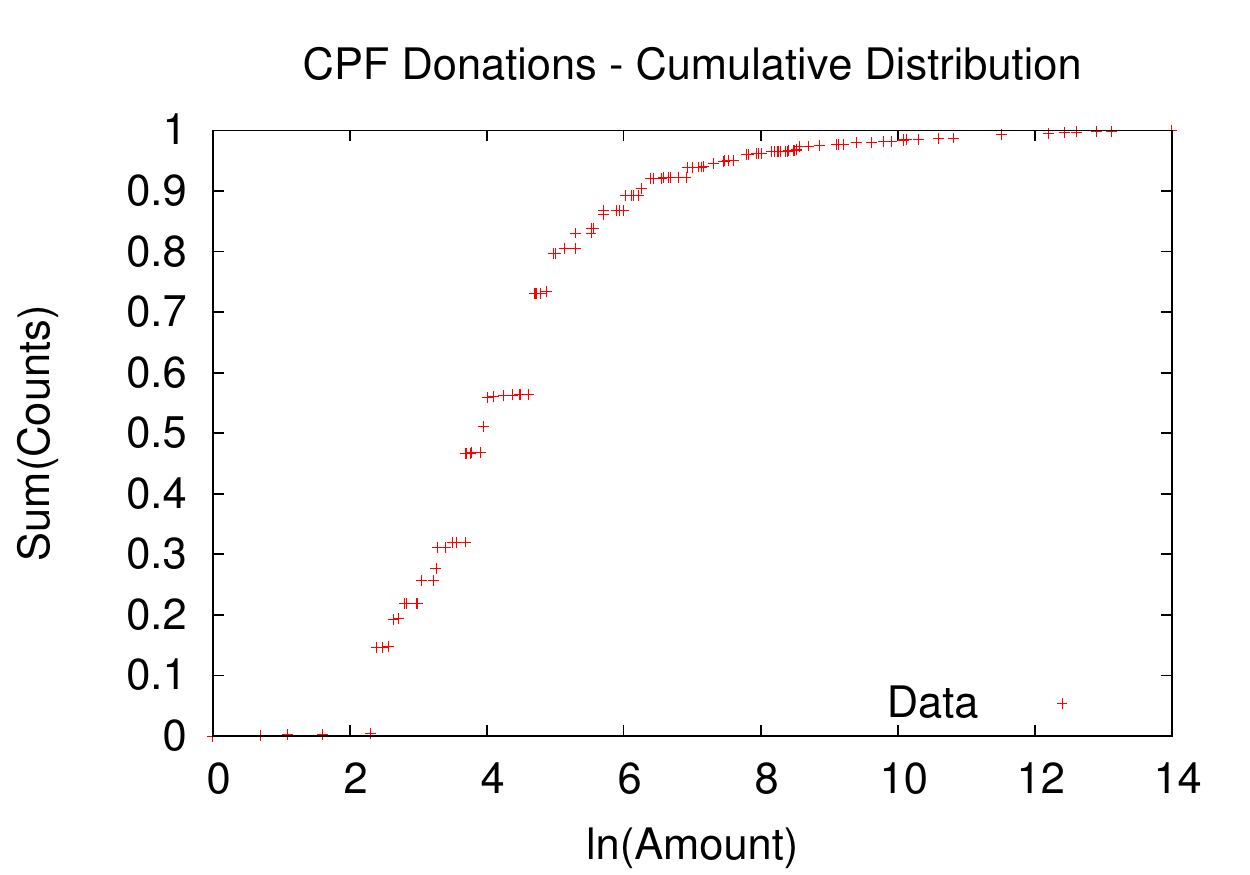}
\et
\ec
\caption{Cumulative distributions for donations declared for the presidential campaigns. Top left: all donations; Top right: non-original donations; Bottom left: CNPJ donations; Bottom right: CPF donations.}\label{fig:allamounts}
\efig

\btab
\caption{Statistics for all donations declared for the presidential campaigns.}\label{tab1}
\bc
\bt{c|cccccc}
Donations & Min [R\$] & Max [R\$] & Average [R\$] & STD [R\$] & N & Total [R\$]  \\
\hline
All &       1.00 & 14000000.00 &  81773.906 & 435812.697 & 11400 & 932222528.31 \\
CNPJ &       1.56 & 14000000.00 & 252883.227 & 874317.765 & 2242 & 566964194.65  \\
CPF &       1.00 & 1200000.00 &   2777.926 &  30722.847 & 5198 & 14439660.54  \\
Non-original &       2.38 & 7000000.00 &  88825.673 & 297193.854 & 3949 & 350772581.57  \\
Unknown &     100.00 &   18826.29 &   4190.141 &   5818.040 & 11 &   46091.55 
\et
\ec
\etab

\btab
\caption{Statistics for all donations declared by the parties central directories.}\label{tab:tabcp}
\bc
\bt{c|cccccc}
Donations & Min [R\$] & Max [R\$] & Average [R\$] & STD [R\$] & N & Total [R\$]  \\
\hline
All &       0.02 & 13000000.00 & 366917.292 & 756161.911 & 2490 & 913624056.53  \\
CNPJ &      30.00 & 13000000.00 & 422592.562 & 802850.503 & 2129 & 899699564.27 \\
CPF &       0.10 & 1000000.00 &  41129.083 & 133472.945 & 336 & 13819371.87  \\
Unknown &       0.02 &   26211.02 &   4204.816 &   6714.171 & 25 &  105120.39  
\et
\ec
\etab

In figure \ref{fig:alldRS} and table \ref{tab2} we show the same histograms and statistics, evaluated for the governor regional campaign in one federal unit, the Rio Grande do Sul (RS). Data for the other 26 federal units, shows very similar patterns. 

\bfig
\bc
\bt{cc}
\ig{.4}{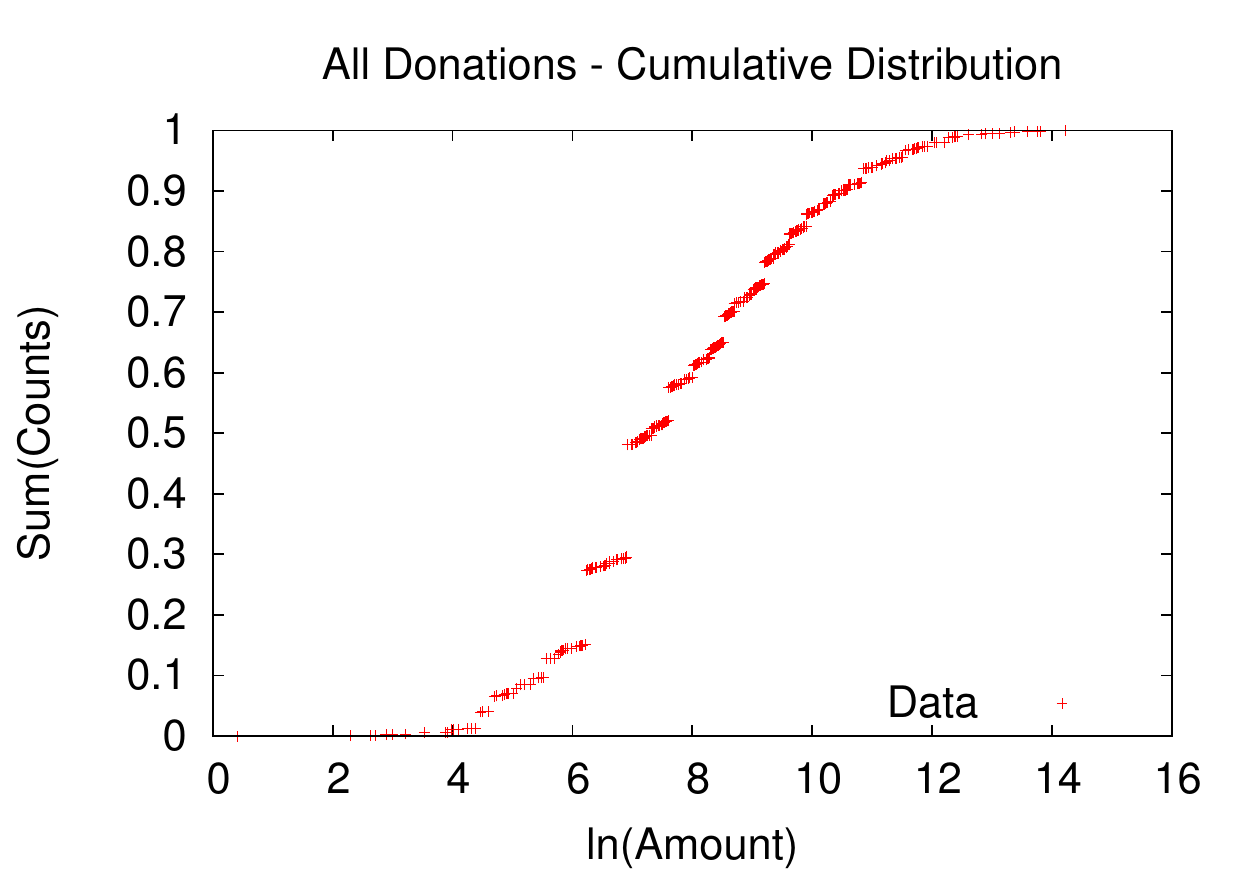} & \ig{.4}{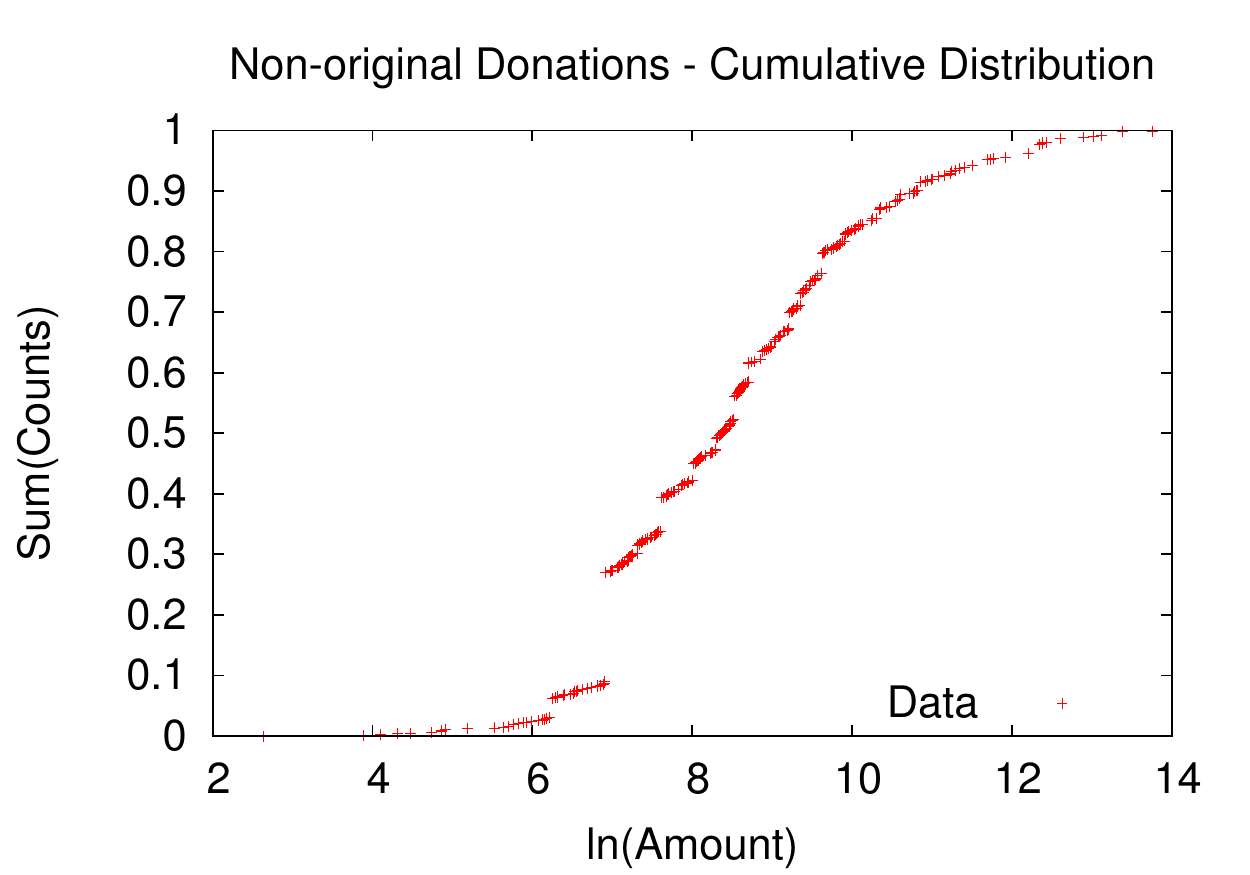} \\
\ig{.4}{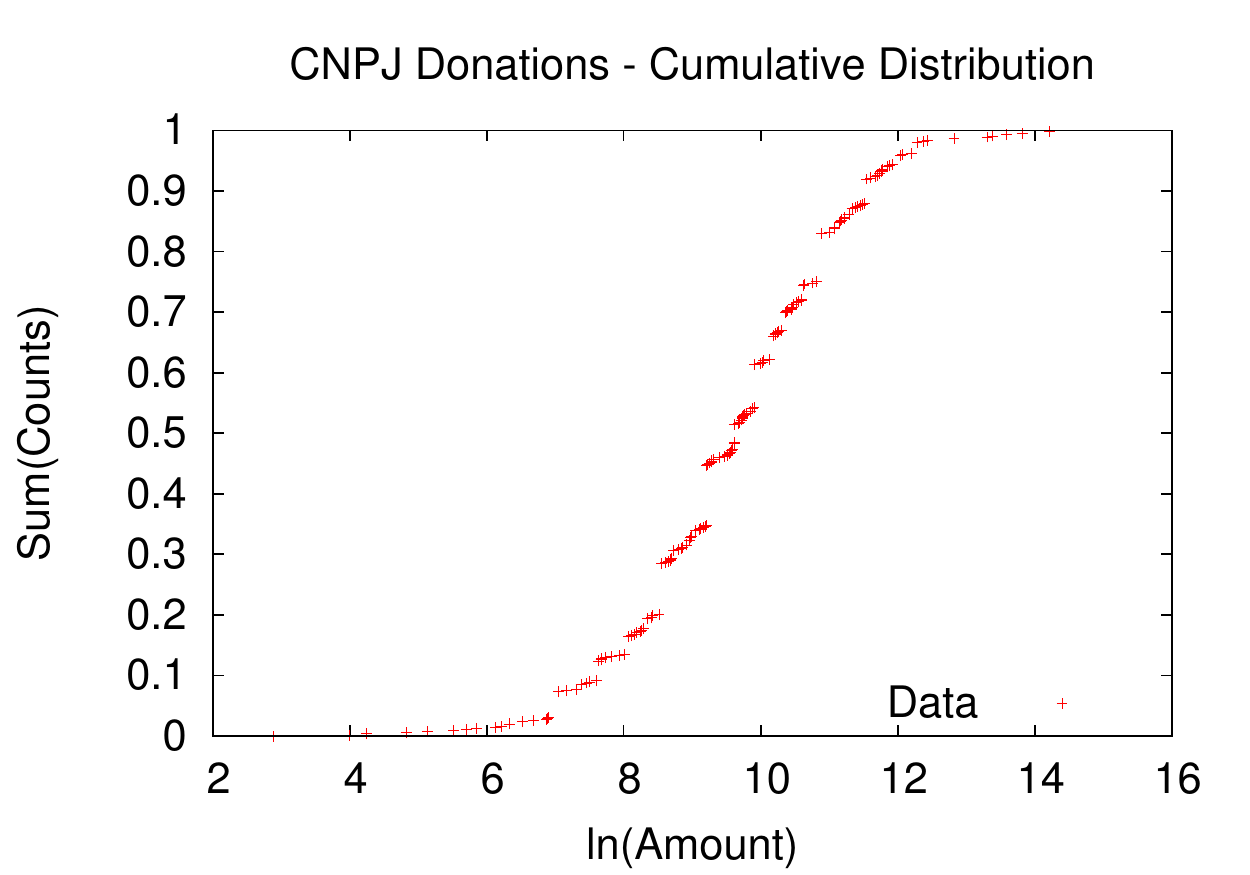} & \ig{.4}{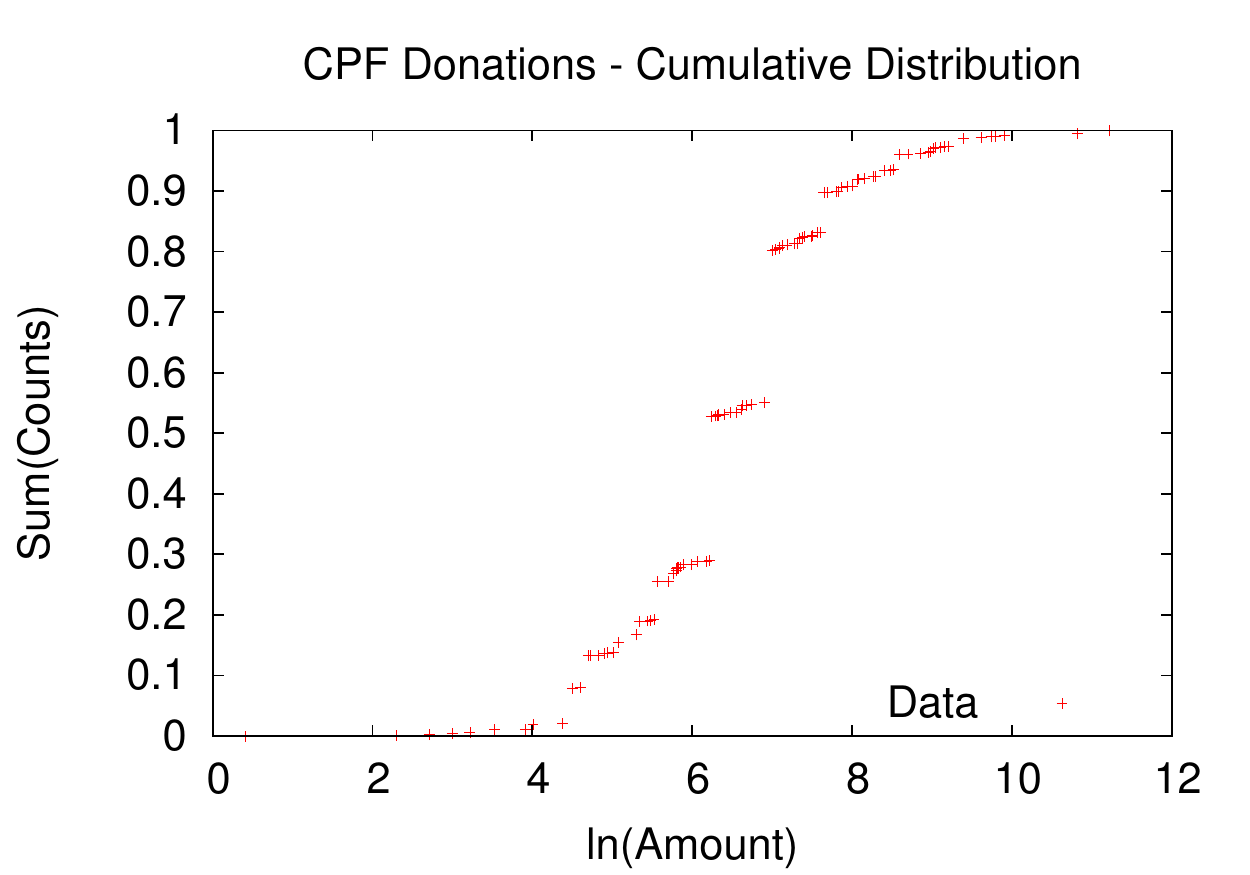}
\et
\ec
\caption{Top: Histogram in logarithmic scale for all donations declared in a regional campaign. Bottom left: only donations from legal entities. Bottom right: only donations from legal persons.}\label{fig:alldRS}
\efig

\btab
\caption{Statistics for all donations declared in the RS Governor electoral campaigns.}\label{tab2}
\bc
\bt{c|ccc|ccc}
Type & Min [R\$] & Max [R\$] &  Average [R\$] & STD [R\$] & N  & Total [R\$] \\
\hline
All &       1.50 & 1500000.00 &  17191.173 &  66018.435 & 2595 & 44611094.90  \\
CNPJ &      18.00 & 1500000.00 &  43165.157 & 108942.678 & 545 & 23525010.55  \\
CPF &       1.50 &   75000.00 &   1523.075 &   4530.274 & 1229 & 1871859.08  \\
Non-original &      14.00 &  950000.00 &  23458.969 &  70986.723 & 819 & 19212895.27  \\
Unknown &     330.00 &    1000.00 &    665.000 &    473.762 & 2 &    1330.00
\et
\ec
\etab

\subsection{Donation Model}

In figures \ref{fig:allamounts} and \ref{fig:alldRS} one can see that the cumulative distribution of the donations tend to have a sigmoidal shape when the horizontal axis is in logarithmic scale. 

Let's fit to the cumulative distributions a truncated sigmoidal function. Take the function

\be
F(\xi) &=& \frac{\xi^\gamma}{\xi_0^\gamma+ \xi^\gamma}=\frac{1}{1+\left(\frac{\xi_0}{\xi}\right)^\gamma} \\
\xi &\in& (0; \infty),
\ee
where $\gamma$ and $\xi_0$ are the parameters to be fitted. In order to shift the minimum possible value of the variable $\xi$, one can make the replacement $\xi\rightarrow \xi-\Delta$ where $\Delta$ is now the new minimum possible value for the variable. In order to truncate the maximum value for the distribution ($F(\xi)$ should be equal to 1 for $\xi_{max}$), one makes the transformation $F(\xi) \rightarrow \frac{F(\xi)}{F(\xi_{max})}$. Since the horizontal axis is in logarithmic scale, the variable of the distribution is related to the amounts ($x$) present in the data by $\xi=\ln(x)$. The actual distribution one needs is the derivative of the cumulative distribution:

\be
F(x) &=& \left\{\ba{cc} 0. & x<e^\Delta \\
                        \frac{1+\left(\frac{\xi_0}{\xi_{max}-\Delta}\right)^\gamma}{1+\left(\frac{\xi_0}{\xi-\Delta}\right)^\gamma} & e^\Delta<x<e^{\xi_{max}} \\
                        1 & x>e^{\xi_{max}} \ea \right. \\ %} 
f(x) &=& \frac{\textrm{d}}{\textrm{d}x}F(x) = \frac{\gamma}{x}\left(1+\left(\frac{\xi_0}{\xi_{max}-\Delta}\right)^\gamma\right)\frac{\left(\frac{\xi_0}{\ln(x)-\Delta}\right)^\gamma}{\left(1+\left(\frac{\xi_0}{\ln(x)-\Delta}\right)^\gamma\right)^2}\frac{1}{\ln(x)-\Delta}. \label{eq:distribution}
\ee

Now, given a set of numbers ($x_i$, $i=1, 2, ..., N$) we set $\xi_{max}=\max(x_i)+1$, $\Delta=log(0.01)$ (the smallest possible donation is one cent) and determine the values of $\xi_0$ and $\gamma$ that maximize the likelihood for the set:

\be
{\cal L} &=& \prod_{i=1}^N f(x_i) \\
\ln{\cal L} &=& \Sum_i f(x_i) = N\left(\ln\gamma+\ln\left(1+\left(\frac{\xi_0}{\xi_{max}-\Delta}\right)^\gamma\right)+\gamma\ln\xi_0\right) - \Sum_{i=1}^N\ln x_i +\nn\\
&-& (\gamma+1)\Sum_{i=1}^N\ln\left(\ln x_i-\Delta\right) - 2\Sum_{i=1}^N\ln\left(1+\left(\frac{\xi_0}{\ln x_i-\Delta}\right)^\gamma\right).
\ee
A steepest ascent algorithm was implemented in order to maximize the likelihood: iteratively, given initial arbitrary values for $\gamma$ and $\xi_0$, the gradient of the likelihood in parameter space is evaluated and the values of the parameters are increased (or decreased) by small amounts following this gradient. This iteration is repeated until the norm of the gradient approaches zero (within some numerical precision). 

In figure \ref{fig:distrfit}, one can see the same distributions from figure \ref{fig:allamounts} with the fitted distributions. The parameters obtained for these fits are in table \ref{tab:fit1}.

\bfig
\bc
\bt{cc}
\ig{.4}{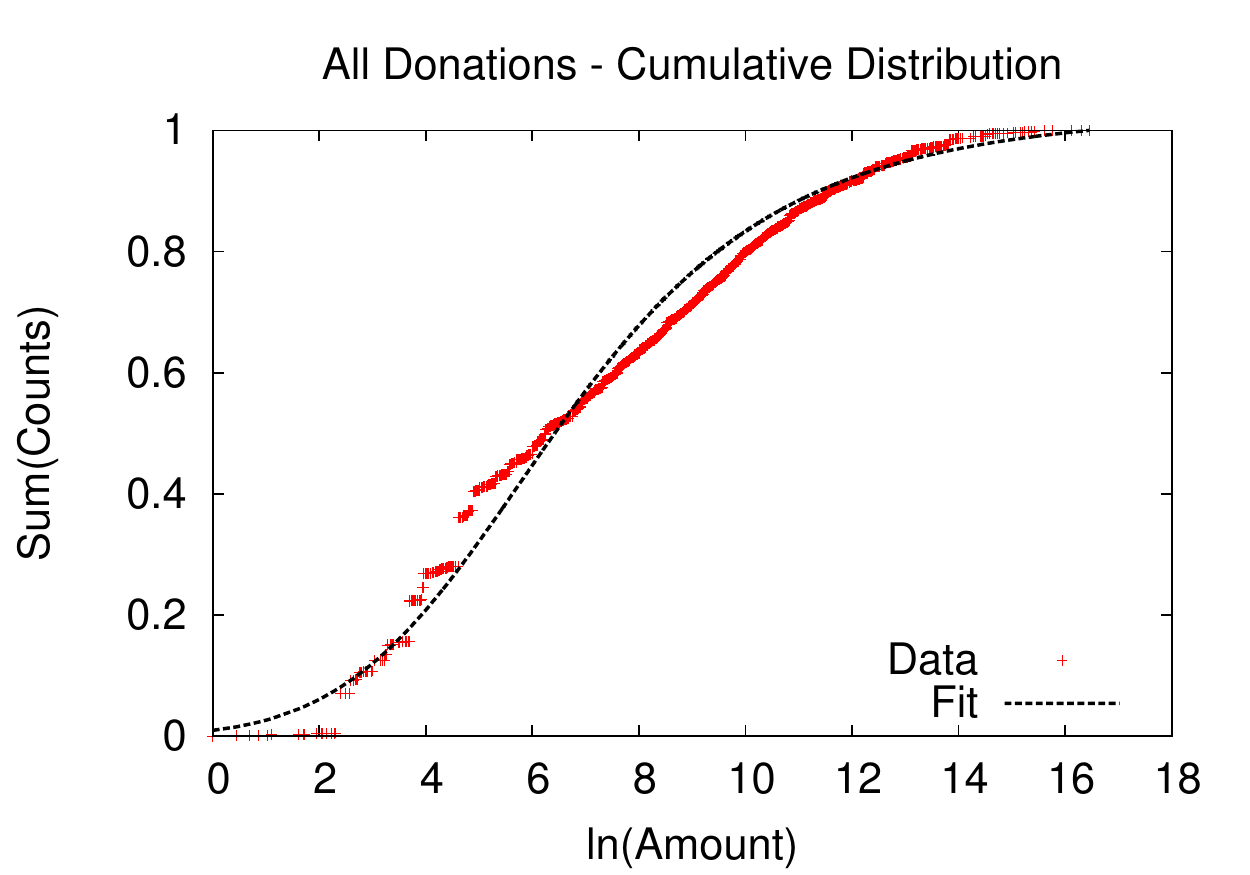} & \ig{.4}{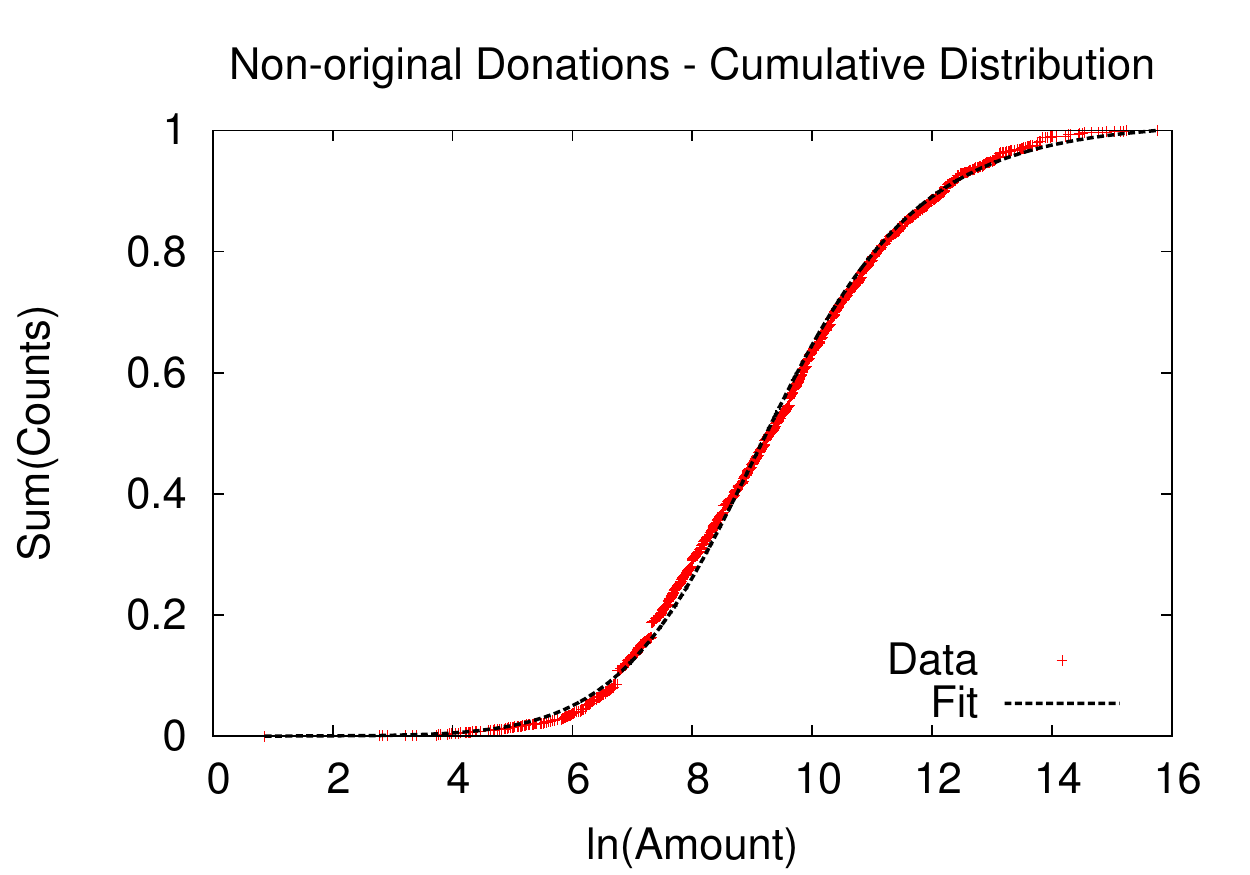} \\
\ig{.4}{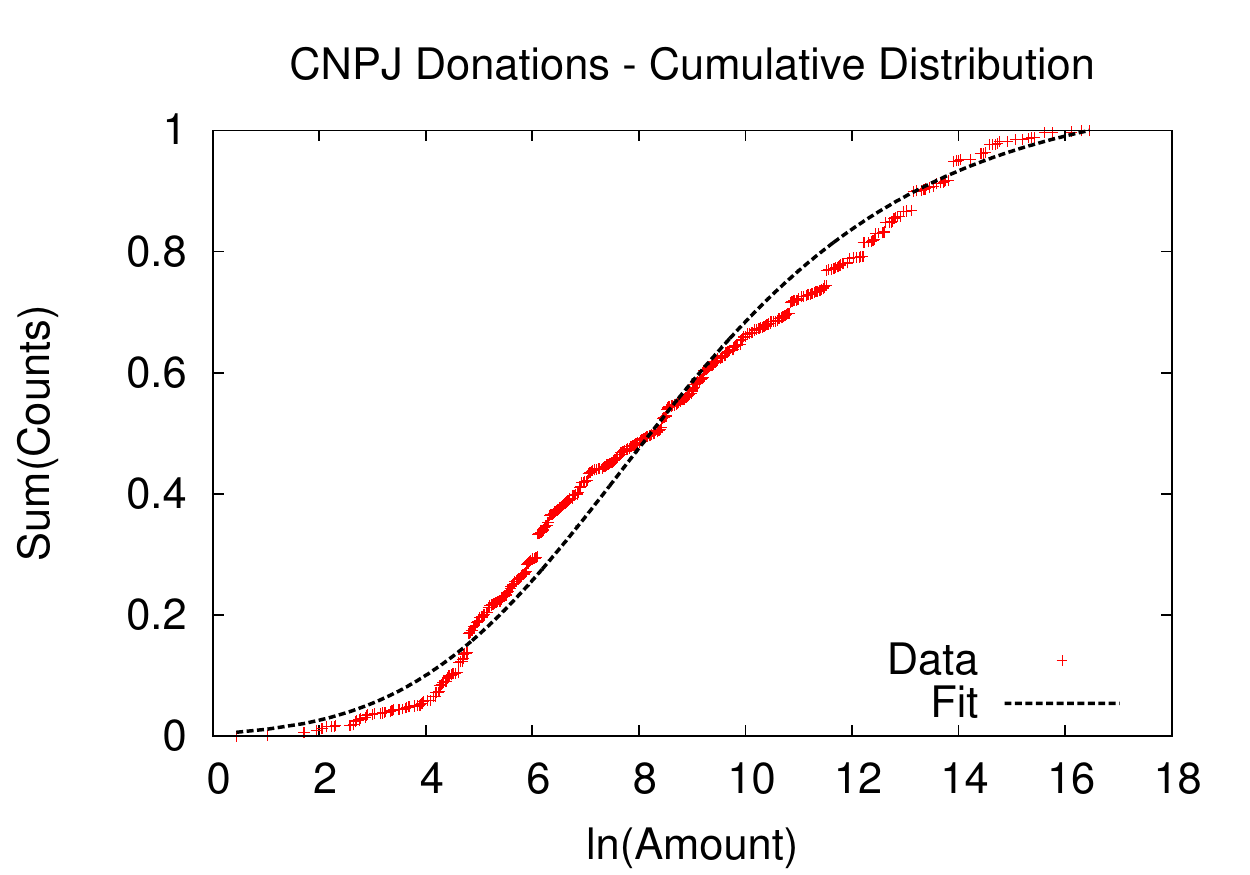} & \ig{.4}{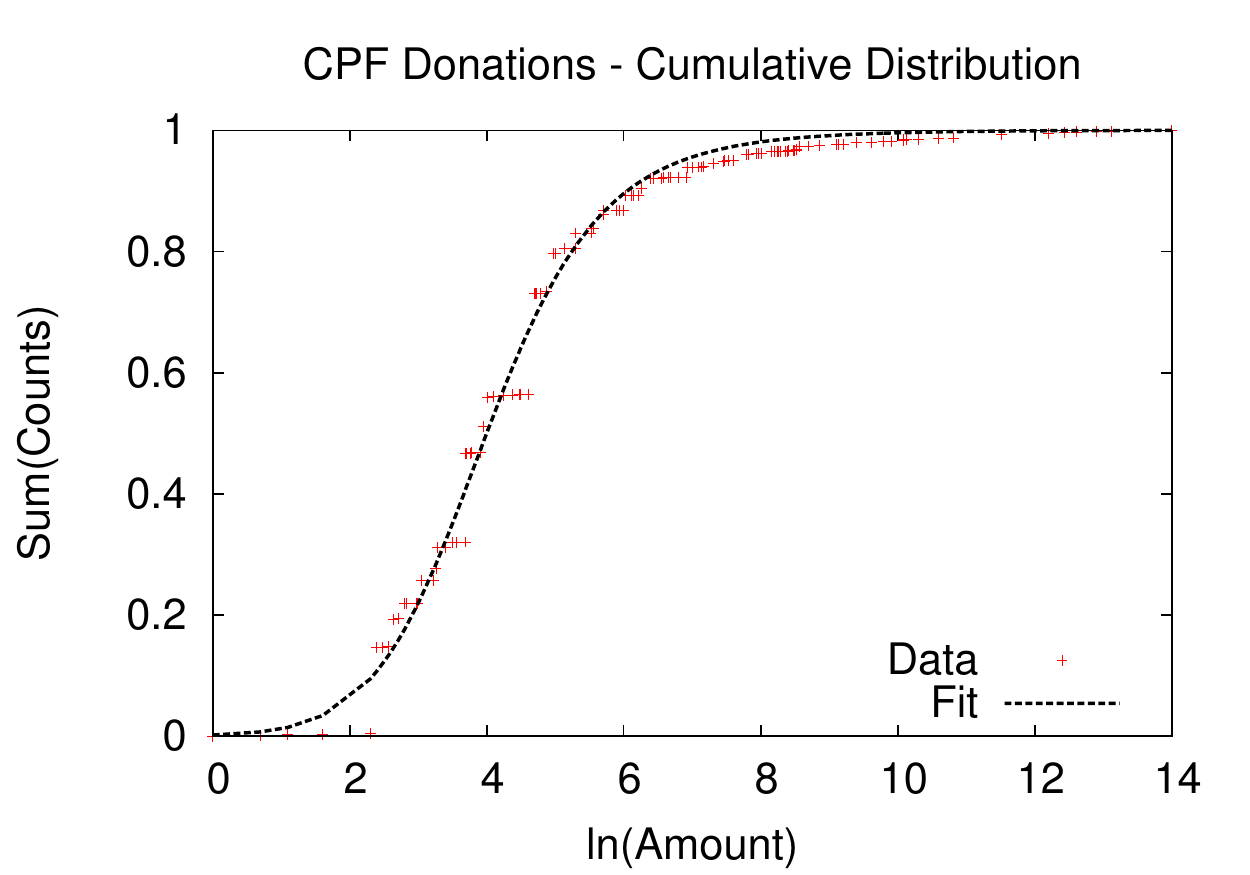}
\et
\ec
\caption{Cumulative distributions for donations declared for the presidential campaigns with fitted distributions. Top left: all donations; Top right: non-original donations; Bottom left: CNPJ donations; Bottom right: CPF donations.}\label{fig:distrfit}
\efig

\btab
\small\caption{Fitted parameters to the distributions in the presidential campaign.}\label{tab:fit1}
\bc
\bt{c|cccc}
Donations & Max [R\$]  & N  & $\gamma$ & $e^{\xi_0}$ [R\$] \\
\hline
All &      14000000.00  & 11400  & 5.289261 &   70747.23 \\
CNPJ &     14000000.00  & 2242  & 5.334680 &  539065.67 \\
CPF &      1200000.00  & 5198  & 10.274493 &    5463.28 \\
Non-original & 7000000.00  & 3949  & 10.982049 & 1068336.77
\et
\ec
\etab

\subsection{Logistic Regression}

Before presenting our results on the first digit distribution for the declared donations in the electoral campaigns, we would like to establish the importance this money has in the outcome of the election. For this purpose, we fit a logistic model (logit regression) setting as dependent variable the result of the election for a given candidate (elected or not elected) and as independent variable the fraction of all money declared as donations in the whole campaign for the same office\footnote{We use the fraction instead of the absolute amount for numerical stability in the calculations. A simple variable transformation, that has no influence in the statistics of the regression results, can be done simply by multiplying the fractions by the total amounts that are shown in the tables.} that each candidate received. 

The logistic model comes from the assumption that one can determine the probability of success ($p$) in a given process from a set of $k$ predictor factors ($x_i$, i=1, 2, ..., $k$) with the logistic function\cite{logit1}:

\be
p(x) &=& \frac{1}{1+e^{-\left(\beta_0+\Sum_{i=1}^k\beta_i x_i\right)}} \label{eq:logit}
\ee
where $p$ is the probability of success, $x_i$ are the values of the predictors and $\beta_i$ are parameters to be fitted from data. In our case, we fit two parameters, $\beta_0$ and $\beta_1$ by correlating the probability of a candidate to be elected for office with the total fraction of donations $x_1$ received by him during the campaign. In order to obtain the best fit to the model we use Newton-Raphson method in order to obtain the values of the parameters that maximize the likelihood for the observed data given the model. 

Regarding the logistic regression, we perform two statistical tests: for each fitted parameter ($\beta_0$ and $\beta_1$), the Wald test where we compare the ratio between the square of the parameter with its uncertainty against a $\chi^2$ distribution with one degree of freedom (this is equivalent to the z-test). Small p-values in this test indicate a significant value for the parameter obtained (that it is in fact different from zero). In another test we evaluate the deviance for the model given the obtained parameters \cite{deviance} and compare it to a $\chi^2$ distribution. This test asserts the overall quality of the fit: high (close to 1) p-values in this test indicate a good fit (a good description of the data by the model).

%%%%%%%%%%%%%%%

\section{Results and Discussion}

For performing the logistic regression analysis, the election results for each one of the 27 Brazilian federal units were obtained. In each federal unit, for each candidate to a given office (we analyze data for the regional and federal congress offices because for these offices many candidates are elected and, therefore, there is enough statistics to perform the analysis) one is able to obtain, from the data, the total amount of money received by him as donations and to know whether he was elected or not. 

In table \ref{tab:logit} we show the results of the logistic regression fit for the federal congressmen campaign in each one of Brazil's federal units. It is astonishing how well the model fits the data (p-values associated to the deviance close to 1) and how significant the obtained parameters are (Wald p-values close to 0). This result indicates that money certainly is a good predictor of whether a candidate is elected or not. From these results, the value of $\beta_1$ can be used to estimate how more likely (how the odds ratio increase) for a candidate to be elected if he receives an extra amount of money $y$:

\be
\textrm{oddR}(y) &=& \frac{p(x+y)}{p(x)} = e^{\beta_1 y}
\ee
take, for example, the case of federal congress campaign in RS, where $\beta_1$=329.49. If a candidate receives a 100000R\$ donation, the odds of him being elected increase $e^{\frac{100000*329.49}{55096519.3}}=1.82$. In other words, the odds of being elected raise by 82\% every 100000R\$ a candidate receives as donation. And receiving nothing, the chances of being elected are $p(0)=0.0133$ (around 1.3\%).

\btab
\caption{Results for the logistic regression fits for the federal congress elections. For each federal unit (UF), the first line show the value for the fitted intercept ($\beta_0$) with its uncertainty and the correspondent Wald p-value and the second line shows the values for the parameter associated with the money variable ($\beta_1$) with its correspondent uncertainty and p-value. The column $N$ shows the number of candidates and the column $n$ the number disputed chairs (elected candidates). The last p-value column is the result for the test performed with the deviance parameter that reflects the overall quality of the fit. In the last column we present the total amount of money donated for the campaigns in each UF.}\label{tab:logit}
\bc
\small
\bt{c|cc|cccc|c}
UF & $\beta\pm\sigma$ & p-value (Wald) & $N$ & $n$ & Deviance & p-value & Total Money [R\$]\\
\hline
RS & -4.301740 $\pm$ 0.485408 & 0.000000 & 308 & 31 & 74.527927 & 1.000000 &  55096519.30\\
   & 329.493809 $\pm$ 52.566813 & 0.000000 &    &    &       &       & \\
\hline
AC & -3.449199 $\pm$ 0.757419 & 0.000005 & 62 & 8 & 32.333029 & 0.998680 &   8413305.29\\
   & 55.166227 $\pm$ 16.037827 & 0.000582 &    &    &       &       & \\
\hline
AL & -4.695802 $\pm$ 0.990996 & 0.000002 & 100 & 9 & 20.127256 & 1.000000 &  17400034.17\\
   & 92.585381 $\pm$ 23.365122 & 0.000074 &    &    &       &       & \\
\hline
AM & -4.707182 $\pm$ 1.167616 & 0.000055 & 79 & 8 & 17.908722 & 1.000000 &  22986481.15\\
   & 91.221557 $\pm$ 26.302040 & 0.000524 &    &    &       &       & \\
\hline
AP & -2.867302 $\pm$ 0.437003 & 0.000000 & 104 & 8 & 49.453531 & 0.999998 &   8668201.78\\
   & 22.829906 $\pm$ 10.653730 & 0.032121 &    &    &       &       & \\
\hline
BA & -3.896939 $\pm$ 0.407161 & 0.000000 & 312 & 39 & 90.934608 & 1.000000 &  71157732.26\\
   & 418.617858 $\pm$ 64.287666 & 0.000000 &    &    &       &       & \\
\hline
CE & -3.503933 $\pm$ 0.433627 & 0.000000 & 195 & 22 & 74.903382 & 1.000000 &  34860757.57\\
   & 137.942303 $\pm$ 23.606311 & 0.000000 &    &    &       &       & \\
\hline
DF & -4.313380 $\pm$ 0.756930 & 0.000000 & 128 & 8 & 36.354865 & 1.000000 &  19218865.40\\
   & 78.962065 $\pm$ 18.805065 & 0.000027 &    &    &       &       & \\
\hline
ES & -4.124638 $\pm$ 0.619728 & 0.000000 & 157 & 10 & 41.305258 & 1.000000 &  18836757.57\\
   & 87.468974 $\pm$ 19.952579 & 0.000012 &    &    &       &       & \\
\hline
GO & -3.070569 $\pm$ 0.534978 & 0.000000 & 95 & 17 & 51.188912 & 0.999870 &  63643888.31\\
   & 93.675674 $\pm$ 20.562509 & 0.000005 &    &    &       &       & \\
\hline
MA & -3.128799 $\pm$ 0.334342 & 0.000000 & 235 & 18 & 102.435737 & 1.000000 &  21227657.48\\
   & 79.044032 $\pm$ 19.063608 & 0.000034 &    &    &       &       & \\
\hline
MG & -4.252349 $\pm$ 0.337550 & 0.000000 & 620 & 53 & 141.959135 & 1.000000 & 158630223.27\\
   & 503.226068 $\pm$ 56.348581 & 0.000000 &    &    &       &       & \\
\hline
MS & -5.973087 $\pm$ 1.580883 & 0.000158 & 116 & 8 & 14.792692 & 1.000000 &  29169885.35\\
   & 128.623347 $\pm$ 35.197929 & 0.000258 &    &    &       &       & \\
\hline
MT & -4.494480 $\pm$ 0.947234 & 0.000002 & 96 & 8 & 21.760442 & 1.000000 &  25722371.33\\
   & 83.391765 $\pm$ 23.946738 & 0.000497 &    &    &       &       & \\
\hline
PA & -3.764366 $\pm$ 0.507372 & 0.000000 & 174 & 17 & 64.372699 & 1.000000 &  19237189.65\\
   & 129.209794 $\pm$ 24.536572 & 0.000000 &    &    &       &       & \\
\hline
PB & -4.795535 $\pm$ 1.083687 & 0.000010 & 96 & 12 & 17.658204 & 1.000000 &  13639599.77\\
   & 157.281575 $\pm$ 44.818445 & 0.000449 &    &    &       &       & \\
\hline
PE & -2.897830 $\pm$ 0.380326 & 0.000000 & 155 & 25 & 85.841997 & 0.999998 &  50791711.79\\
   & 126.721817 $\pm$ 23.957652 & 0.000000 &    &    &       &       & \\
\hline
PI & -3.362935 $\pm$ 0.626560 & 0.000000 & 88 & 10 & 38.131270 & 0.999998 &  24434149.38\\
   & 80.046119 $\pm$ 23.186551 & 0.000556 &    &    &       &       & \\
\hline
PR & -3.522870 $\pm$ 0.348306 & 0.000000 & 295 & 30 & 99.829366 & 1.000000 &  69943697.86\\
   & 198.013071 $\pm$ 31.148521 & 0.000000 &    &    &       &       & \\
\hline
RJ & -4.309115 $\pm$ 0.278953 & 0.000000 & 953 & 46 & 182.068165 & 1.000000 & 109193461.16\\
   & 497.533828 $\pm$ 56.288402 & 0.000000 &    &    &       &       & \\
\hline
RN & -5.692015 $\pm$ 1.756848 & 0.001196 & 83 & 8 & 13.736874 & 1.000000 &  14278235.86\\
   & 110.857108 $\pm$ 35.288882 & 0.001681 &    &    &       &       & \\
\hline
RO & -3.420208 $\pm$ 0.646803 & 0.000000 & 81 & 8 & 35.803050 & 0.999993 &  17001787.46\\
   & 56.756842 $\pm$ 18.389173 & 0.002026 &    &    &       &       & \\
\hline
RR & -4.050292 $\pm$ 0.846561 & 0.000002 & 80 & 8 & 25.829202 & 1.000000 &   8356682.91\\
   & 96.267954 $\pm$ 29.868455 & 0.001268 &    &    &       &       & \\
\hline
SC & -3.715785 $\pm$ 0.579054 & 0.000000 & 128 & 16 & 49.832673 & 1.000000 &  31737303.92\\
   & 115.394467 $\pm$ 23.029840 & 0.000001 &    &    &       &       & \\
\hline
SE & -2.981286 $\pm$ 0.564645 & 0.000000 & 73 & 8 & 36.331160 & 0.999795 &   7622064.24\\
   & 37.830953 $\pm$ 13.895874 & 0.006480 &    &    &       &       & \\
\hline
SP & -4.113540 $\pm$ 0.215463 & 0.000000 & 1318 & 70 & 301.879103 & 1.000000 & 234478433.27\\
   & 551.467211 $\pm$ 47.421220 & 0.000000 &    &    &       &       & \\
\hline
TO & -3.353321 $\pm$ 0.853338 & 0.000085 & 47 & 8 & 25.137784 & 0.992752 &  14895843.70\\
   & 53.979352 $\pm$ 17.658840 & 0.002237 &    &    &       &       & \\
\hline
\et
\ec
\etab

In Table \ref{tab:logit_est} we show the results of the logistic regression for the regional congress office seats.

\btab[H]
\caption{Results for the logistic regression fits for the regional congress elections. For each federal unit (UF), the first line show the value for the fitted intercept ($\beta_0$) with its uncertainty and the correspondent Wald p-value and the second line shows the values for the parameter associated with the money variable ($\beta_1$) with its correspondent uncertainty and p-value. The column $N$ shows the number of candidates and the column $n$ the number of disputed chairs (elected candidates). The last p-value column is the result for the test performed with the deviance parameter that reflects the overall quality of the fit. In the last column we present the total amount of money donated for the campaigns in each UF.}\label{tab:logit_est}
\bc
\small
\bt{c|cc|cccc|c}
UF & $\beta\pm\sigma$ & p-value (Wald) & $N$ & $n$ & Deviance & p-value & Total Money [R\$]\\
\hline
RS & -3.895895 $\pm$ 0.268780 & 0.000000 & 670 & 55 & 229.721642 & 1.000000 &  53897470.35\\
   & 520.063531 $\pm$ 57.562793 & 0.000000 &    &    &       &       & \\
\hline
AC & -4.268271 $\pm$ 0.368601 & 0.000000 & 497 & 24 & 122.215355 & 1.000000 &  10741882.23\\
   & 281.612445 $\pm$ 42.665487 & 0.000000 &    &    &       &       & \\
\hline
AL & -3.855591 $\pm$ 0.433952 & 0.000000 & 262 & 27 & 93.457895 & 1.000000 &  19512010.69\\
   & 211.488057 $\pm$ 31.624984 & 0.000000 &    &    &       &       & \\
\hline
AM & -5.421684 $\pm$ 0.560394 & 0.000000 & 570 & 24 & 76.805367 & 1.000000 &  24576432.18\\
   & 416.471023 $\pm$ 60.800589 & 0.000000 &    &    &       &       & \\
\hline
AP & -3.838433 $\pm$ 0.373426 & 0.000000 & 338 & 24 & 112.245178 & 1.000000 &   5621758.62\\
   & 206.887652 $\pm$ 34.012981 & 0.000000 &    &    &       &       & \\
\hline
BA & -3.591355 $\pm$ 0.260449 & 0.000000 & 578 & 63 & 235.229572 & 1.000000 &  47098049.37\\
   & 488.121206 $\pm$ 54.578145 & 0.000000 &    &    &       &       & \\
\hline
CE & -3.866844 $\pm$ 0.296504 & 0.000000 & 558 & 46 & 177.125300 & 1.000000 &  32583088.46\\
   & 368.870924 $\pm$ 42.414498 & 0.000000 &    &    &       &       & \\
\hline
DF & -4.473329 $\pm$ 0.292542 & 0.000000 & 979 & 24 & 158.940820 & 1.000000 &  35648405.99\\
   & 261.270656 $\pm$ 40.316546 & 0.000000 &    &    &       &       & \\
\hline
ES & -4.086863 $\pm$ 0.350796 & 0.000000 & 472 & 30 & 138.579334 & 1.000000 &  23230725.98\\
   & 285.100562 $\pm$ 37.708774 & 0.000000 &    &    &       &       & \\
\hline
GO & -3.736298 $\pm$ 0.245030 & 0.000000 & 717 & 41 & 211.477478 & 1.000000 &  79100331.98\\
   & 270.582549 $\pm$ 32.741606 & 0.000000 &    &    &       &       & \\
\hline
MA & -4.304746 $\pm$ 0.386074 & 0.000000 & 490 & 42 & 119.106959 & 1.000000 &  25769715.01\\
   & 470.819026 $\pm$ 60.737138 & 0.000000 &    &    &       &       & \\
\hline
MG & -3.812026 $\pm$ 0.209294 & 0.000000 & 1055 & 77 & 325.294508 & 1.000000 & 141181882.89\\
   & 571.563491 $\pm$ 49.725496 & 0.000000 &    &    &       &       & \\
\hline
MS & -5.025201 $\pm$ 0.569786 & 0.000000 & 395 & 24 & 69.193927 & 1.000000 &  45931385.85\\
   & 338.506140 $\pm$ 51.048838 & 0.000000 &    &    &       &       & \\
\hline
MT & -3.863263 $\pm$ 0.404347 & 0.000000 & 292 & 24 & 100.184791 & 1.000000 &  51160968.29\\
   & 199.266811 $\pm$ 30.810483 & 0.000000 &    &    &       &       & \\
\hline
PA & -3.599082 $\pm$ 0.243462 & 0.000000 & 650 & 41 & 223.961945 & 1.000000 &  31697472.64\\
   & 276.652126 $\pm$ 36.713893 & 0.000000 &    &    &       &       & \\
\hline
PB & -4.058142 $\pm$ 0.418954 & 0.000000 & 333 & 36 & 110.491532 & 1.000000 &  17238137.47\\
   & 315.883453 $\pm$ 40.915450 & 0.000000 &    &    &       &       & \\
\hline
PE & -3.656680 $\pm$ 0.288765 & 0.000000 & 489 & 49 & 175.615591 & 1.000000 &  40611046.64\\
   & 342.514907 $\pm$ 39.141519 & 0.000000 &    &    &       &       & \\
\hline
PI & -5.604635 $\pm$ 0.884008 & 0.000000 & 226 & 30 & 57.843668 & 1.000000 &  20247713.90\\
   & 412.376534 $\pm$ 69.711243 & 0.000000 &    &    &       &       & \\
\hline
PR & -4.621801 $\pm$ 0.348345 & 0.000000 & 738 & 54 & 166.557103 & 1.000000 &  61112695.05\\
   & 692.209375 $\pm$ 71.533492 & 0.000000 &    &    &       &       & \\
\hline
RJ & -4.316389 $\pm$ 0.199879 & 0.000000 & 1846 & 70 & 342.421533 & 1.000000 & 129527220.71\\
   & 630.840620 $\pm$ 54.192459 & 0.000000 &    &    &       &       & \\
\hline
RN & -4.152926 $\pm$ 0.509765 & 0.000000 & 244 & 24 & 69.343989 & 1.000000 &  18313268.55\\
   & 198.547307 $\pm$ 30.987689 & 0.000000 &    &    &       &       & \\
\hline
RO & -3.658539 $\pm$ 0.325359 & 0.000000 & 382 & 24 & 130.082195 & 1.000000 &  25257051.30\\
   & 176.974288 $\pm$ 30.220810 & 0.000000 &    &    &       &       & \\
\hline
RR & -4.162472 $\pm$ 0.394217 & 0.000000 & 393 & 24 & 100.233152 & 1.000000 &  13405608.74\\
   & 227.670731 $\pm$ 32.618995 & 0.000000 &    &    &       &       & \\
\hline
SC & -3.662375 $\pm$ 0.315007 & 0.000000 & 409 & 40 & 159.849890 & 1.000000 &  51935831.01\\
   & 323.898860 $\pm$ 43.323434 & 0.000000 &    &    &       &       & \\
\hline
SE & -3.379372 $\pm$ 0.456329 & 0.000000 & 162 & 24 & 83.869900 & 1.000000 &   8427039.38\\
   & 161.551846 $\pm$ 28.397770 & 0.000000 &    &    &       &       & \\
\hline
SP & -4.193506 $\pm$ 0.185004 & 0.000000 & 1878 & 94 & 432.635429 & 1.000000 & 229364470.60\\
   & 836.922136 $\pm$ 60.696497 & 0.000000 &    &    &       &       & \\
\hline
TO & -4.478863 $\pm$ 0.587435 & 0.000000 & 238 & 24 & 74.147005 & 1.000000 &  19956645.63\\
   & 290.774731 $\pm$ 50.526507 & 0.000000 &    &    &       &       & \\
\hline
\et
\ec
\etab

The results clearly indicate that money is an excellent predictor of whether a given candidate will be elected for office or not. Note, that the model deals with probabilities, so we are not saying that the ones who receive more money are surely elected; what the analysis shows is that the more money a candidate receives, the more probable it is for him to be elected. This might (sadly) sound like an obvious statement, but this actually obliterates the principle of democracy: our representatives are not elected because they represent our interests, they are elected because they gather huge amounts of money in order to hire marketing professionals and flood media time with spurious publicity. In other words, candidates know that, to get elected, they do not need to have previously done a good job, to have developed good projects or even to be honest, the most important thing they need is to get money, lots of it.

It could be possible try and argue that the candidates that receive the most money are those that better represent the population, since it is the population that makes the donations. Actually, most of the money comes from legal entities (CNPJ) or from the parties (non-original) which recieved most of their money from companies. Moreover, it is common knowledge that most parties and candidates receive unofficial donations (in Portuguese referred to by the term ``{\it caixa dois}'', second cashier) that must pass through some kind of money laundry before being used. It is quite easy to compare the huge amounts donated by companies with plain bribes: the candidates, once elected, legislate beneficial laws to help those economic sectors that helped them to be elected in the first place; from the companies point of view, the donations are investments \cite{financing}.

The next analysis is based on the idea that the set of numbers in an honest financial declaration should follow the Benford's law for the first significant digit distribution. Some arguments for that have already been presented: the declarator should have no real control over the amounts (random donations) and the amounts expand many orders of magnitude and come from multiple different distributions. Therefore, deviations of Benford's law would be evidence of fraud, indications that the declaration has been ``cooked'' (money laundry). 

In table \ref{tab:benfordBR} we show the first significant digit distribution for all donations received by each political party specifically for the 2014 presidential campaign. For each party, we present the results for all donations together and then for the amounts classified according to the donor (CNPJ, CPF, non-original). We also present results for artificial data generated from the distributions fitted to each data set; those are indicated by the tag \texttt{\_Rand} and in the last line, for each party, indicated by the tag \texttt{\_Model}, is the result for the different sets of artificial data for each category combined. In this analysis, we have considered only sets of data with more than 20 elements (some parties or some specific categories in a given party that are not shown in the table, had $N<20$ and were omitted). In table \ref{tab:benfordCP} we present a similar table for all the donations declared by each political party central directory. These are the donations that end up distributed as non-original to various campaigns. 

The $\chi^2$ values in the table are calculated via:

\be
\chi^2 &=& \sum_{i=1}^9 N\frac{(O_i-E_i)^2}{E_i}
\ee
where $O_i$ is the observed frequency of each digit, $E_i$ is the expected frequency according to Benford's Law and $N$ is the total number of elements in the analyzed set. The p-values are the probabilities that a fluctuation equal or bigger than the observed $\chi^2$ may be obtained in a set of numbers of the same size that do agree with the Benford empirical distribution.

\begin{sidewaystable}
\btab[H]
\caption{First significant digit distribution for all donations received for the presidential campaign by the different parties. The columns 1-9 indicate the proportion each one of the nine digits is observed, N the total number of donations, $\chi^2$ is the statistic for the fluctuation, p-val its correspondent p-value, Min and Max are respectively the minimum and maximum amounts donated and sum is the sum of all donations, $\gamma$ and $\xi_0$ are the model parameters.}\label{tab:benfordBR}
\tiny
\bc
\bt{c|ccccccccc|ccc|ccccc}
Partido & 1 & 2 & 3 & 4 & 5 & 6 & 7 & 8 & 9 & N & $\chi^2$ & p-val & Min & Max & Sum & $\gamma$ & $\xi_0$\\
\hline
\hline
Benford & 0.301 & 0.176 & 0.125 & 0.097 & 0.079 & 0.067 & 0.058 & 0.051 & 0.046 & & & & &\\
\hline
PSDB - All & 0.304 & 0.183 & 0.115 & 0.091 & 0.088 & 0.060 & 0.049 & 0.068 & 0.043 & 3545 & 36.968 & 0.000 &    17.00 & 14000000.00 & 428137499.87 & 10.6744 & 13.9580 \\
PSDB - All Rand & 0.292 & 0.181 & 0.129 & 0.092 & 0.084 & 0.065 & 0.054 & 0.049 & 0.054 & 3545 & 10.023 & 0.263 &    10.58 & 13553439.45 & 647168618.07 & 10.6744 & 13.9580 \\
PSDB - CNPJ & 0.263 & 0.222 & 0.080 & 0.086 & 0.198 & 0.047 & 0.027 & 0.038 & 0.038 & 338 & 81.172 & 0.000 &    20.00 & 14000000.00 & 176814668.46 & 9.9850 & 16.4027 \\
PSDB - CNPJ Rand & 0.243 & 0.180 & 0.151 & 0.080 & 0.077 & 0.053 & 0.071 & 0.080 & 0.065 & 338 & 16.872 & 0.031 &   127.87 & 13656868.89 & 290580085.73 & 9.9850 & 16.4027 \\
PSDB - CPF & 0.268 & 0.390 & 0.098 & 0.049 & 0.122 & 0.016 & 0.049 & 0.000 & 0.008 & 123 & 53.990 & 0.000 &   450.00 & 1200000.00 & 7933331.08 & 11.4890 & 13.3403 \\
PSDB - CPF Rand & 0.285 & 0.228 & 0.138 & 0.073 & 0.057 & 0.049 & 0.041 & 0.049 & 0.081 & 123 & 8.280 & 0.407 &   105.78 & 428568.20 & 4612216.49 & 11.4890 & 13.3403 \\
PSDB - Non-original & 0.310 & 0.169 & 0.120 & 0.093 & 0.074 & 0.063 & 0.051 & 0.074 & 0.045 & 3078 & 39.221 & 0.000 &    17.00 & 7000000.00 & 243369609.95 & 11.1858 & 13.7907 \\
PSDB - Non-original Rand & 0.305 & 0.187 & 0.126 & 0.090 & 0.083 & 0.069 & 0.048 & 0.048 & 0.045 & 3078 & 9.838 & 0.277 &     3.77 & 6998795.84 & 322481463.76 & 11.1858 & 13.7907 \\
PSDB - Model & 0.298 & 0.188 & 0.129 & 0.089 & 0.081 & 0.067 & 0.050 & 0.051 & 0.048 & 3545 & 9.759 & 0.282 &     3.77 & 13656868.89 & 622410419.10 & - & - \\
\hline
PV - All & 0.267 & 0.206 & 0.170 & 0.079 & 0.109 & 0.042 & 0.067 & 0.030 & 0.030 & 165 & 10.517 & 0.231 &     1.56 & 1000000.00 & 7922280.69 & 5.9339 & 12.0622 \\
PV - All Rand & 0.352 & 0.152 & 0.145 & 0.055 & 0.103 & 0.048 & 0.061 & 0.042 & 0.042 & 165 & 7.905 & 0.443 &     0.48 & 979516.62 & 5255748.70 & 5.9339 & 12.0622 \\
PV - CNPJ & 0.242 & 0.226 & 0.169 & 0.089 & 0.056 & 0.056 & 0.081 & 0.040 & 0.040 & 124 & 7.698 & 0.464 &     1.56 & 1000000.00 & 7893364.47 & 5.5143 & 13.2271 \\
PV - CNPJ Rand & 0.339 & 0.153 & 0.113 & 0.105 & 0.081 & 0.073 & 0.065 & 0.024 & 0.048 & 124 & 3.111 & 0.927 &     0.95 & 935712.73 & 10042984.38 & 5.5143 & 13.2271 \\
PV - CPF & 0.341 & 0.146 & 0.171 & 0.049 & 0.268 & 0.000 & 0.024 & 0.000 & 0.000 & 41 & 28.131 & 0.000 &    10.00 &  7000.00 & 28916.22 & 9.9028 & 10.0454 \\
PV - CPF Rand & 0.268 & 0.317 & 0.049 & 0.073 & 0.049 & 0.122 & 0.049 & 0.073 & 0.000 & 41 & 11.572 & 0.171 &     6.36 &  5294.41 & 28733.08 & 9.9028 & 10.0454 \\
PV - Model & 0.321 & 0.194 & 0.097 & 0.097 & 0.073 & 0.085 & 0.061 & 0.036 & 0.036 & 165 & 3.475 & 0.901 &     0.95 & 935712.73 & 10071717.47 & - & - \\
\hline
PSTU - All & 0.496 & 0.141 & 0.133 & 0.067 & 0.081 & 0.030 & 0.015 & 0.022 & 0.015 & 135 & 31.598 & 0.000 &    20.00 & 20000.00 & 171676.75 & 9.7162 & 10.1468 \\
PSTU - All Rand & 0.281 & 0.178 & 0.148 & 0.081 & 0.104 & 0.067 & 0.052 & 0.037 & 0.052 & 135 & 2.836 & 0.944 &     3.61 & 15837.56 & 109612.51 & 9.7162 & 10.1468 \\
PSTU - CPF & 0.546 & 0.102 & 0.139 & 0.074 & 0.083 & 0.028 & 0.009 & 0.000 & 0.019 & 108 & 39.909 & 0.000 &    20.00 & 20000.00 & 104045.00 & 11.1369 & 9.7449 \\
PSTU - CPF Rand & 0.306 & 0.222 & 0.176 & 0.074 & 0.046 & 0.056 & 0.019 & 0.056 & 0.046 & 108 & 8.769 & 0.362 &     2.16 &  5493.99 & 46241.56 & 11.1369 & 9.7449 \\
PSTU - Model & 0.356 & 0.200 & 0.148 & 0.067 & 0.044 & 0.044 & 0.030 & 0.059 & 0.052 & 135 & 8.862 & 0.354 &     2.16 & 19258.32 & 320209.16 & - & - \\
\hline
PT - All & 0.411 & 0.182 & 0.059 & 0.071 & 0.170 & 0.022 & 0.034 & 0.028 & 0.022 & 3024 & 751.337 & 0.000 &     1.00 & 10000000.00 & 351270140.26 & 6.1549 & 10.4056 \\
PT - All Rand & 0.303 & 0.182 & 0.126 & 0.091 & 0.080 & 0.070 & 0.063 & 0.045 & 0.040 & 3024 & 8.129 & 0.421 &     0.11 & 9993290.98 & 241002479.22 & 6.1549 & 10.4056 \\
PT - CNPJ & 0.365 & 0.133 & 0.090 & 0.129 & 0.099 & 0.040 & 0.061 & 0.053 & 0.031 & 1478 & 96.414 & 0.000 &     2.80 & 10000000.00 & 307721104.56 & 5.8880 & 11.8043 \\
PT - CNPJ Rand & 0.317 & 0.172 & 0.136 & 0.085 & 0.077 & 0.062 & 0.066 & 0.046 & 0.039 & 1478 & 9.835 & 0.277 &     0.20 & 9919764.38 & 245410372.53 & 5.8880 & 11.8043 \\
PT - CPF & 0.485 & 0.235 & 0.007 & 0.004 & 0.264 & 0.002 & 0.002 & 0.001 & 0.002 & 1320 & 1290.454 & 0.000 &     1.00 & 500000.00 & 869017.00 & 11.3187 & 8.8406 \\
PT - CPF Rand & 0.278 & 0.181 & 0.137 & 0.091 & 0.076 & 0.069 & 0.067 & 0.053 & 0.048 & 1320 & 6.855 & 0.552 &     1.01 & 62840.98 & 466397.20 & 11.3187 & 8.8406 \\
PT - Non-original & 0.283 & 0.195 & 0.159 & 0.088 & 0.084 & 0.031 & 0.044 & 0.027 & 0.088 & 226 & 19.852 & 0.011 &    16.20 & 3800000.00 & 42680018.70 & 10.3475 & 14.7289 \\
PT - Non-original Rand & 0.367 & 0.133 & 0.128 & 0.088 & 0.088 & 0.040 & 0.071 & 0.035 & 0.049 & 226 & 10.399 & 0.238 &    15.32 & 3722448.46 & 35609479.64 & 10.3475 & 14.7289 \\
PT - Model & 0.304 & 0.173 & 0.136 & 0.088 & 0.077 & 0.063 & 0.067 & 0.048 & 0.044 & 3024 & 11.365 & 0.182 &     0.20 & 9919764.38 & 281486249.37 & - & - \\
\hline
PSB - All & 0.472 & 0.132 & 0.040 & 0.244 & 0.077 & 0.011 & 0.014 & 0.005 & 0.006 & 3876 & 2104.931 & 0.000 &     1.00 & 5000000.00 & 123023025.46 & 6.6782 & 9.0822 \\
PSB - All Rand & 0.303 & 0.174 & 0.117 & 0.102 & 0.079 & 0.068 & 0.061 & 0.049 & 0.047 & 3876 & 4.537 & 0.806 &     0.13 & 4708204.53 & 55901564.13 & 6.6782 & 9.0822 \\
PSB - CNPJ & 0.353 & 0.199 & 0.096 & 0.038 & 0.212 & 0.032 & 0.045 & 0.000 & 0.026 & 156 & 55.538 & 0.000 &    13.33 & 5000000.00 & 54326607.36 & 5.8168 & 17.5564 \\
PSB - CNPJ Rand & 0.301 & 0.103 & 0.154 & 0.115 & 0.071 & 0.071 & 0.071 & 0.051 & 0.064 & 156 & 8.129 & 0.421 &     6.25 & 4697620.88 & 80401952.53 & 5.8168 & 17.5564 \\
PSB - CPF & 0.508 & 0.118 & 0.018 & 0.282 & 0.065 & 0.004 & 0.003 & 0.002 & 0.001 & 3142 & 2560.399 & 0.000 &     1.00 & 500000.00 & 5387549.24 & 11.1146 & 8.4009 \\
PSB - CPF Rand & 0.281 & 0.194 & 0.128 & 0.101 & 0.079 & 0.067 & 0.056 & 0.050 & 0.043 & 3142 & 11.228 & 0.189 &     0.15 & 304842.86 & 1492020.92 & 11.1146 & 8.4009 \\
PSB - Non-original & 0.307 & 0.186 & 0.141 & 0.085 & 0.106 & 0.047 & 0.068 & 0.026 & 0.033 & 574 & 21.131 & 0.007 &     2.38 & 3000000.00 & 63301493.98 & 11.2112 & 14.3195 \\
PSB - Non-original Rand & 0.287 & 0.179 & 0.157 & 0.111 & 0.068 & 0.075 & 0.045 & 0.040 & 0.037 & 574 & 11.799 & 0.160 &    26.64 & 2661538.45 & 67001138.08 & 11.2112 & 14.3195 \\
PSB - Model & 0.282 & 0.188 & 0.133 & 0.103 & 0.077 & 0.069 & 0.055 & 0.049 & 0.043 & 3876 & 13.331 & 0.101 &     0.15 & 4697620.88 & 148977209.29 & - & - \\
\hline
PDT - All & 0.291 & 0.236 & 0.164 & 0.091 & 0.073 & 0.055 & 0.036 & 0.018 & 0.036 & 55 & 3.707 & 0.883 &  4000.00 & 2000000.00 & 18938826.29 & 15.0178 & 16.6327 \\
PDT - All Rand & 0.236 & 0.200 & 0.127 & 0.036 & 0.109 & 0.109 & 0.055 & 0.055 & 0.073 & 55 & 6.004 & 0.647 &  1210.40 & 1913866.96 & 13948573.89 & 15.0178 & 16.6327 \\
PDT - CNPJ & 0.278 & 0.241 & 0.167 & 0.093 & 0.074 & 0.056 & 0.037 & 0.019 & 0.037 & 54 & 3.887 & 0.867 &  4000.00 & 2000000.00 & 18920000.00 & 15.1149 & 16.6790 \\
PDT - CNPJ Rand & 0.241 & 0.130 & 0.093 & 0.130 & 0.074 & 0.093 & 0.093 & 0.074 & 0.074 & 54 & 5.527 & 0.700 &   835.23 & 1822964.94 & 18306741.04 & 15.1149 & 16.6790 \\
PDT - Model & 0.255 & 0.127 & 0.091 & 0.127 & 0.073 & 0.091 & 0.091 & 0.073 & 0.073 & 55 & 5.075 & 0.750 &   835.23 & 1822964.94 & 19328360.54 & - & - \\
\hline
PRTB - All & 0.328 & 0.138 & 0.155 & 0.069 & 0.069 & 0.086 & 0.000 & 0.000 & 0.155 & 58 & 23.410 & 0.003 &    43.30 & 100000.00 & 440985.26 & 11.1681 & 11.5622 \\
PRTB - All Rand & 0.190 & 0.138 & 0.172 & 0.103 & 0.121 & 0.017 & 0.155 & 0.034 & 0.069 & 58 & 17.787 & 0.023 &    16.16 & 54157.04 & 246427.41 & 11.1681 & 11.5622 \\
PRTB - Non-original & 0.368 & 0.105 & 0.132 & 0.079 & 0.079 & 0.105 & 0.000 & 0.000 & 0.132 & 38 & 12.893 & 0.116 &    43.30 & 100000.00 & 195090.32 & 12.5753 & 11.3702 \\
PRTB - Non-original Rand & 0.289 & 0.368 & 0.079 & 0.053 & 0.105 & 0.026 & 0.000 & 0.079 & 0.000 & 38 & 15.191 & 0.056 &    29.20 & 87262.96 & 367477.89 & 12.5753 & 11.3702 \\
PRTB - Model & 0.259 & 0.310 & 0.052 & 0.052 & 0.086 & 0.052 & 0.086 & 0.069 & 0.034 & 58 & 11.548 & 0.173 &    29.20 & 90014.86 & 1238572.93 & - & - \\
\hline
PCB - All & 0.273 & 0.152 & 0.121 & 0.152 & 0.000 & 0.152 & 0.061 & 0.061 & 0.030 & 33 & 7.592 & 0.474 &    70.00 & 16000.00 & 60554.69 & 13.5250 & 11.1803 \\
PCB - All Rand & 0.303 & 0.212 & 0.121 & 0.061 & 0.030 & 0.182 & 0.030 & 0.000 & 0.061 & 33 & 10.480 & 0.233 &    54.45 &  3125.84 & 35234.71 & 13.5250 & 11.1803 \\
\hline
PSOL - All & 0.437 & 0.265 & 0.028 & 0.006 & 0.251 & 0.002 & 0.004 & 0.004 & 0.002 & 471 & 393.503 & 0.000 &     1.00 & 94953.74 & 401555.67 & 9.8255 & 8.4428 \\
PSOL - All Rand & 0.291 & 0.195 & 0.123 & 0.072 & 0.079 & 0.064 & 0.062 & 0.053 & 0.062 & 471 & 6.923 & 0.545 &     1.50 & 14644.12 & 129531.55 & 9.8255 & 8.4428 \\
PSOL - CNPJ & 0.167 & 0.417 & 0.000 & 0.042 & 0.208 & 0.000 & 0.083 & 0.042 & 0.042 & 24 & 20.061 & 0.010 &    18.40 & 94953.74 & 246919.27 & 9.8356 & 13.0237 \\
PSOL - CNPJ Rand & 0.542 & 0.167 & 0.042 & 0.083 & 0.042 & 0.000 & 0.083 & 0.000 & 0.042 & 24 & 9.542 & 0.299 &    56.21 & 43474.71 & 261107.25 & 9.8356 & 13.0237 \\
PSOL - CPF & 0.451 & 0.260 & 0.027 & 0.005 & 0.255 & 0.002 & 0.000 & 0.000 & 0.000 & 443 & 393.235 & 0.000 &     1.00 & 12000.00 & 103538.00 & 11.1878 & 8.3139 \\
PSOL - CPF Rand & 0.300 & 0.187 & 0.120 & 0.088 & 0.081 & 0.050 & 0.070 & 0.052 & 0.052 & 443 & 4.251 & 0.834 &     0.25 &  8307.16 & 71625.52 & 11.1878 & 8.3139 \\
PSOL - Model & 0.316 & 0.185 & 0.115 & 0.087 & 0.079 & 0.047 & 0.070 & 0.049 & 0.053 & 471 & 6.106 & 0.635 &     0.25 & 43474.71 & 363857.42 & - & - \\
\hline
\et
\ec
\etab
\end{sidewaystable}

\begin{sidewaystable}
\btab[H]
\caption{First significant digit distribution for all donations received for the presidential campaign by the different parties. The columns 1-9 indicate the proportion each one of the nine digits is observed, N the total number of donations, $\chi^2$ is the statistic for the fluctuation, p-val its correspondent p-value, Min and Max are respectively the minimum and maximum amounts donated and sum is the sum of all donations.}\label{tab:benfordCP}
\tiny
\bc
\bt{c|ccccccccc|ccc|ccc}
Partido & 1 & 2 & 3 & 4 & 5 & 6 & 7 & 8 & 9 & N & $\chi^2$ & p-val & Min & Max & Sum \\
\hline
\hline
Benford & 0.301 & 0.176 & 0.125 & 0.097 & 0.079 & 0.067 & 0.058 & 0.051 & 0.046 & & & & &\\
\hline
PT do B - All & 0.417 & 0.000 & 0.083 & 0.000 & 0.333 & 0.000 & 0.167 & 0.000 & 0.000 & 12 & 18.175 & 0.020 &  1200.00 & 105000.00 & 408402.23 \\
PT do B - CNPJ & 0.417 & 0.000 & 0.083 & 0.000 & 0.333 & 0.000 & 0.167 & 0.000 & 0.000 & 12 & 18.175 & 0.020 &  1200.00 & 105000.00 & 408402.23 \\
\hline
PEN - All & 0.667 & 0.000 & 0.000 & 0.000 & 0.333 & 0.000 & 0.000 & 0.000 & 0.000 & 3 & 5.639 & 0.688 &  1500.00 & 500000.00 & 503000.00 \\
PEN - CNPJ & 0.000 & 0.000 & 0.000 & 0.000 & 1.000 & 0.000 & 0.000 & 0.000 & 0.000 & 1 & 11.629 & 0.169 & 500000.00 & 500000.00 & 500000.00 \\
PEN - CPF & 1.000 & 0.000 & 0.000 & 0.000 & 0.000 & 0.000 & 0.000 & 0.000 & 0.000 & 2 & 4.644 & 0.795 &  1500.00 &  1500.00 &  3000.00 \\
\hline
DEM - All & 0.364 & 0.193 & 0.107 & 0.079 & 0.179 & 0.021 & 0.021 & 0.021 & 0.014 & 140 & 33.400 & 0.000 &   600.00 & 3300000.00 & 51921770.00 \\
DEM - CNPJ & 0.350 & 0.197 & 0.109 & 0.080 & 0.182 & 0.022 & 0.022 & 0.022 & 0.015 & 137 & 32.995 & 0.000 &   600.00 & 3300000.00 & 51621770.00 \\
DEM - CPF & 1.000 & 0.000 & 0.000 & 0.000 & 0.000 & 0.000 & 0.000 & 0.000 & 0.000 & 3 & 6.966 & 0.540 & 100000.00 & 100000.00 & 300000.00 \\
\hline
PMDB - All & 0.302 & 0.181 & 0.129 & 0.055 & 0.189 & 0.031 & 0.058 & 0.024 & 0.031 & 381 & 79.460 & 0.000 &  1513.58 & 7000000.00 & 186799588.38 \\
PMDB - CNPJ & 0.318 & 0.188 & 0.128 & 0.045 & 0.176 & 0.031 & 0.060 & 0.026 & 0.028 & 352 & 65.570 & 0.000 &  3000.95 & 7000000.00 & 182368186.61 \\
PMDB - CPF & 0.103 & 0.103 & 0.138 & 0.172 & 0.345 & 0.034 & 0.034 & 0.000 & 0.069 & 29 & 34.778 & 0.000 &  1513.58 & 1000000.00 & 4431401.77 \\
\hline
PR - All & 0.339 & 0.271 & 0.059 & 0.085 & 0.203 & 0.025 & 0.008 & 0.000 & 0.008 & 118 & 51.511 & 0.000 &   542.39 & 5000000.00 & 63275909.49 \\
PR - CNPJ & 0.339 & 0.284 & 0.064 & 0.083 & 0.193 & 0.018 & 0.009 & 0.000 & 0.009 & 109 & 46.059 & 0.000 &  2500.00 & 5000000.00 & 62324300.02 \\
PR - CPF & 0.333 & 0.167 & 0.000 & 0.000 & 0.333 & 0.167 & 0.000 & 0.000 & 0.000 & 6 & 8.070 & 0.427 &  5000.00 & 650000.00 & 930000.00 \\
\hline
PTN - All & 0.292 & 0.167 & 0.042 & 0.125 & 0.083 & 0.000 & 0.250 & 0.042 & 0.000 & 24 & 19.556 & 0.012 &    21.38 & 800000.00 & 5659820.94 \\
PTN - CNPJ & 0.250 & 0.062 & 0.062 & 0.188 & 0.062 & 0.000 & 0.375 & 0.000 & 0.000 & 16 & 33.569 & 0.000 & 50000.00 & 731000.00 & 4854350.00 \\
PTN - CPF & 0.667 & 0.000 & 0.000 & 0.000 & 0.000 & 0.000 & 0.000 & 0.333 & 0.000 & 3 & 7.946 & 0.439 &   100.00 & 800000.00 & 800200.00 \\
\hline
PP - All & 0.266 & 0.210 & 0.153 & 0.048 & 0.218 & 0.024 & 0.040 & 0.024 & 0.016 & 124 & 43.363 & 0.000 & 10000.00 & 13000000.00 & 72497495.13 \\
PP - CNPJ & 0.254 & 0.213 & 0.156 & 0.049 & 0.221 & 0.025 & 0.041 & 0.025 & 0.016 & 122 & 44.621 & 0.000 & 10000.00 & 13000000.00 & 72297495.13 \\
PP - CPF & 1.000 & 0.000 & 0.000 & 0.000 & 0.000 & 0.000 & 0.000 & 0.000 & 0.000 & 2 & 4.644 & 0.795 & 100000.00 & 100000.00 & 200000.00 \\
\hline
PV - All & 0.354 & 0.228 & 0.114 & 0.114 & 0.076 & 0.038 & 0.025 & 0.025 & 0.025 & 79 & 6.471 & 0.595 &    10.00 & 1000000.00 & 7909375.23 \\
PV - CNPJ & 0.346 & 0.231 & 0.115 & 0.115 & 0.077 & 0.038 & 0.026 & 0.026 & 0.026 & 78 & 6.224 & 0.622 &   228.62 & 1000000.00 & 7909365.23 \\
PV - CPF & 1.000 & 0.000 & 0.000 & 0.000 & 0.000 & 0.000 & 0.000 & 0.000 & 0.000 & 1 & 2.322 & 0.970 &    10.00 &    10.00 &    10.00 \\
\hline
PT - All & 0.379 & 0.177 & 0.127 & 0.042 & 0.169 & 0.023 & 0.044 & 0.023 & 0.016 & 385 & 84.742 & 0.000 &    10.00 & 10000000.00 & 191128848.27 \\
PT - CNPJ & 0.378 & 0.182 & 0.133 & 0.044 & 0.155 & 0.022 & 0.047 & 0.025 & 0.014 & 362 & 68.508 & 0.000 &   326.26 & 10000000.00 & 188973838.27 \\
PT - CPF & 0.391 & 0.087 & 0.043 & 0.000 & 0.391 & 0.043 & 0.000 & 0.000 & 0.043 & 23 & 36.111 & 0.000 &    10.00 & 500000.00 & 2155010.00 \\
\hline
PRB - All & 0.500 & 0.158 & 0.158 & 0.040 & 0.119 & 0.010 & 0.010 & 0.005 & 0.000 & 202 & 75.135 & 0.000 &     0.02 & 1000000.00 & 10518326.73 \\
PRB - CNPJ & 0.319 & 0.250 & 0.111 & 0.083 & 0.181 & 0.014 & 0.028 & 0.014 & 0.000 & 72 & 21.317 & 0.006 &  4000.00 & 1000000.00 & 10506015.10 \\
PRB - CPF & 0.619 & 0.095 & 0.183 & 0.008 & 0.087 & 0.008 & 0.000 & 0.000 & 0.000 & 126 & 86.824 & 0.000 &    10.00 &   500.00 &  4865.00 \\
\hline
PTC - All & 0.000 & 0.000 & 0.000 & 0.500 & 0.500 & 0.000 & 0.000 & 0.000 & 0.000 & 2 & 9.474 & 0.304 & 50000.00 & 400000.00 & 450000.00 \\
PTC - CNPJ & 0.000 & 0.000 & 0.000 & 0.500 & 0.500 & 0.000 & 0.000 & 0.000 & 0.000 & 2 & 9.474 & 0.304 & 50000.00 & 400000.00 & 450000.00 \\
\hline
PTB - All & 0.267 & 0.333 & 0.167 & 0.033 & 0.167 & 0.000 & 0.033 & 0.000 & 0.000 & 30 & 14.129 & 0.078 & 10000.00 & 1300000.00 & 8810000.00 \\
PTB - CNPJ & 0.267 & 0.333 & 0.167 & 0.033 & 0.167 & 0.000 & 0.033 & 0.000 & 0.000 & 30 & 14.129 & 0.078 & 10000.00 & 1300000.00 & 8810000.00 \\
\hline
PPL - All & 0.143 & 0.143 & 0.429 & 0.000 & 0.143 & 0.000 & 0.000 & 0.000 & 0.143 & 7 & 9.503 & 0.302 &    30.12 & 333403.93 & 439890.53 \\
PPL - CNPJ & 0.250 & 0.000 & 0.500 & 0.000 & 0.000 & 0.000 & 0.000 & 0.000 & 0.250 & 4 & 10.298 & 0.245 &    90.00 & 333403.93 & 437360.41 \\
PPL - CPF & 0.000 & 0.500 & 0.000 & 0.000 & 0.500 & 0.000 & 0.000 & 0.000 & 0.000 & 2 & 7.154 & 0.520 &   500.00 &  2000.00 &  2500.00 \\
\hline
PPS - All & 0.320 & 0.200 & 0.120 & 0.040 & 0.240 & 0.000 & 0.040 & 0.000 & 0.040 & 25 & 12.227 & 0.141 &   500.15 & 500000.00 & 2598945.87 \\
PPS - CNPJ & 0.263 & 0.158 & 0.158 & 0.053 & 0.263 & 0.000 & 0.053 & 0.000 & 0.053 & 19 & 11.071 & 0.198 & 30000.00 & 500000.00 & 2545000.00 \\
PPS - CPF & 0.500 & 0.500 & 0.000 & 0.000 & 0.000 & 0.000 & 0.000 & 0.000 & 0.000 & 4 & 5.001 & 0.757 &  1500.00 & 20000.00 & 51500.00 \\
\hline
PRP - All & 0.000 & 0.500 & 0.500 & 0.000 & 0.000 & 0.000 & 0.000 & 0.000 & 0.000 & 2 & 4.841 & 0.774 &  2500.00 & 300000.00 & 302500.00 \\
PRP - CNPJ & 0.000 & 0.000 & 1.000 & 0.000 & 0.000 & 0.000 & 0.000 & 0.000 & 0.000 & 1 & 7.004 & 0.536 & 300000.00 & 300000.00 & 300000.00 \\
PRP - CPF & 0.000 & 1.000 & 0.000 & 0.000 & 0.000 & 0.000 & 0.000 & 0.000 & 0.000 & 1 & 4.679 & 0.791 &  2500.00 &  2500.00 &  2500.00 \\
\hline
PMN - All & 0.133 & 0.233 & 0.267 & 0.167 & 0.067 & 0.000 & 0.033 & 0.033 & 0.067 & 30 & 12.546 & 0.128 &    30.00 & 1300000.00 & 2485542.30 \\
PMN - CNPJ & 0.148 & 0.259 & 0.296 & 0.074 & 0.074 & 0.000 & 0.037 & 0.037 & 0.074 & 27 & 12.247 & 0.141 &    30.00 & 1300000.00 & 2472242.30 \\
PMN - CPF & 0.000 & 0.000 & 0.000 & 1.000 & 0.000 & 0.000 & 0.000 & 0.000 & 0.000 & 3 & 27.957 & 0.000 &  4100.00 &  4700.00 & 13300.00 \\
\hline
PSDC - All & 0.214 & 0.286 & 0.071 & 0.000 & 0.214 & 0.143 & 0.000 & 0.071 & 0.000 & 14 & 8.980 & 0.344 &   185.96 & 100000.00 & 165885.96 \\
PSDC - CNPJ & 0.231 & 0.308 & 0.077 & 0.000 & 0.154 & 0.154 & 0.000 & 0.077 & 0.000 & 13 & 6.891 & 0.548 &   185.96 & 100000.00 & 165385.96 \\
PSDC - CPF & 0.000 & 0.000 & 0.000 & 0.000 & 1.000 & 0.000 & 0.000 & 0.000 & 0.000 & 1 & 11.629 & 0.169 &   500.00 &   500.00 &   500.00 \\
\hline
PSDB - All & 0.322 & 0.221 & 0.109 & 0.057 & 0.169 & 0.044 & 0.044 & 0.025 & 0.008 & 366 & 69.617 & 0.000 &     0.10 & 9000000.00 & 156718898.33 \\
PSDB - CNPJ & 0.307 & 0.226 & 0.120 & 0.060 & 0.166 & 0.039 & 0.048 & 0.024 & 0.009 & 332 & 59.653 & 0.000 &   935.06 & 9000000.00 & 152093150.51 \\
PSDB - CPF & 0.536 & 0.143 & 0.000 & 0.036 & 0.214 & 0.036 & 0.000 & 0.036 & 0.000 & 28 & 19.777 & 0.011 &     0.10 & 1000000.00 & 4561050.10 \\
\hline
PDT - All & 0.111 & 0.222 & 0.667 & 0.000 & 0.000 & 0.000 & 0.000 & 0.000 & 0.000 & 9 & 25.909 & 0.001 & 30000.00 & 350000.00 & 1050000.00 \\
PDT - CNPJ & 0.111 & 0.222 & 0.667 & 0.000 & 0.000 & 0.000 & 0.000 & 0.000 & 0.000 & 9 & 25.909 & 0.001 & 30000.00 & 350000.00 & 1050000.00 \\
\hline
PRTB - All & 0.211 & 0.053 & 0.158 & 0.105 & 0.263 & 0.158 & 0.053 & 0.000 & 0.000 & 19 & 14.661 & 0.066 &    50.00 & 31200.00 & 156631.94 \\
PRTB - CNPJ & 0.222 & 0.056 & 0.167 & 0.111 & 0.278 & 0.167 & 0.000 & 0.000 & 0.000 & 18 & 16.573 & 0.035 &    50.00 & 31200.00 & 155907.94 \\
PRTB - CPF & 0.000 & 0.000 & 0.000 & 0.000 & 0.000 & 0.000 & 1.000 & 0.000 & 0.000 & 1 & 16.244 & 0.039 &   724.00 &   724.00 &   724.00 \\
\hline
PSL - All & 0.340 & 0.191 & 0.149 & 0.043 & 0.170 & 0.064 & 0.000 & 0.021 & 0.021 & 47 & 11.042 & 0.199 &  1500.00 & 400000.00 & 1777990.00 \\
PSL - CNPJ & 0.340 & 0.191 & 0.149 & 0.043 & 0.170 & 0.064 & 0.000 & 0.021 & 0.021 & 47 & 11.042 & 0.199 &  1500.00 & 400000.00 & 1777990.00 \\
\hline
PSTU - All & 0.400 & 0.200 & 0.000 & 0.000 & 0.000 & 0.400 & 0.000 & 0.000 & 0.000 & 5 & 10.743 & 0.217 &  1986.00 & 18000.00 & 35031.08 \\
PSTU - CNPJ & 0.333 & 0.000 & 0.000 & 0.000 & 0.000 & 0.667 & 0.000 & 0.000 & 0.000 & 3 & 18.024 & 0.021 &  6000.00 & 18000.00 & 30995.08 \\
PSTU - CPF & 0.500 & 0.500 & 0.000 & 0.000 & 0.000 & 0.000 & 0.000 & 0.000 & 0.000 & 2 & 2.500 & 0.962 &  1986.00 &  2050.00 &  4036.00 \\
\hline
PSB - All & 0.190 & 0.171 & 0.124 & 0.076 & 0.324 & 0.038 & 0.029 & 0.019 & 0.029 & 105 & 89.765 & 0.000 &  6270.32 & 3000000.00 & 35986511.21 \\
PSB - CNPJ & 0.192 & 0.173 & 0.125 & 0.077 & 0.327 & 0.038 & 0.029 & 0.019 & 0.019 & 104 & 91.587 & 0.000 &  6270.32 & 3000000.00 & 35896511.21 \\
PSB - CPF & 0.000 & 0.000 & 0.000 & 0.000 & 0.000 & 0.000 & 0.000 & 0.000 & 1.000 & 1 & 20.854 & 0.008 & 90000.00 & 90000.00 & 90000.00 \\
\hline
PSC - All & 0.304 & 0.152 & 0.109 & 0.043 & 0.304 & 0.022 & 0.022 & 0.000 & 0.043 & 46 & 35.862 & 0.000 &   150.00 & 1050000.00 & 5906900.99 \\
PSC - CNPJ & 0.268 & 0.146 & 0.122 & 0.024 & 0.341 & 0.024 & 0.024 & 0.000 & 0.049 & 41 & 42.213 & 0.000 & 10000.00 & 1050000.00 & 5899650.99 \\
PSC - CPF & 0.600 & 0.200 & 0.000 & 0.200 & 0.000 & 0.000 & 0.000 & 0.000 & 0.000 & 5 & 4.179 & 0.841 &   150.00 &  4900.00 &  7250.00 \\
\hline
PCO - All & 0.500 & 0.000 & 0.000 & 0.000 & 0.000 & 0.000 & 0.000 & 0.500 & 0.000 & 2 & 9.436 & 0.307 &    83.33 &   100.00 &   183.33 \\
PCO - CNPJ & 0.500 & 0.000 & 0.000 & 0.000 & 0.000 & 0.000 & 0.000 & 0.500 & 0.000 & 2 & 9.436 & 0.307 &    83.33 &   100.00 &   183.33 \\
\hline
PROS - All & 0.000 & 0.000 & 0.500 & 0.000 & 0.500 & 0.000 & 0.000 & 0.000 & 0.000 & 6 & 24.950 & 0.002 &  3000.00 & 3000000.00 & 4803000.00 \\
PROS - CNPJ & 0.000 & 0.000 & 0.400 & 0.000 & 0.600 & 0.000 & 0.000 & 0.000 & 0.000 & 5 & 24.136 & 0.002 & 300000.00 & 3000000.00 & 4800000.00 \\
PROS - CPF & 0.000 & 0.000 & 1.000 & 0.000 & 0.000 & 0.000 & 0.000 & 0.000 & 0.000 & 1 & 7.004 & 0.536 &  3000.00 &  3000.00 &  3000.00 \\
\hline
PSD - All & 0.348 & 0.197 & 0.182 & 0.045 & 0.121 & 0.015 & 0.076 & 0.000 & 0.015 & 66 & 13.373 & 0.100 & 30000.00 & 5000000.00 & 55736308.58 \\
PSD - CNPJ & 0.354 & 0.200 & 0.185 & 0.046 & 0.108 & 0.015 & 0.077 & 0.000 & 0.015 & 65 & 12.680 & 0.123 & 30000.00 & 5000000.00 & 55686308.58 \\
PSD - CPF & 0.000 & 0.000 & 0.000 & 0.000 & 1.000 & 0.000 & 0.000 & 0.000 & 0.000 & 1 & 11.629 & 0.169 & 50000.00 & 50000.00 & 50000.00 \\
\hline
PC do B - All & 0.333 & 0.242 & 0.152 & 0.030 & 0.121 & 0.091 & 0.000 & 0.030 & 0.000 & 33 & 7.360 & 0.498 &  2500.00 & 5000000.00 & 17275014.99 \\
PC do B - CNPJ & 0.355 & 0.194 & 0.161 & 0.032 & 0.129 & 0.097 & 0.000 & 0.032 & 0.000 & 31 & 6.834 & 0.555 & 20000.00 & 5000000.00 & 17270014.99 \\
PC do B - CPF & 0.000 & 1.000 & 0.000 & 0.000 & 0.000 & 0.000 & 0.000 & 0.000 & 0.000 & 2 & 9.358 & 0.313 &  2500.00 &  2500.00 &  5000.00 \\
\hline
PHS - All & 0.000 & 0.000 & 1.000 & 0.000 & 0.000 & 0.000 & 0.000 & 0.000 & 0.000 & 1 & 7.004 & 0.536 &  3000.00 &  3000.00 &  3000.00 \\
PHS - CPF & 0.000 & 0.000 & 1.000 & 0.000 & 0.000 & 0.000 & 0.000 & 0.000 & 0.000 & 1 & 7.004 & 0.536 &  3000.00 &  3000.00 &  3000.00 \\
\hline
PSOL - All & 0.341 & 0.114 & 0.187 & 0.008 & 0.228 & 0.024 & 0.024 & 0.065 & 0.008 & 123 & 61.401 & 0.000 &    10.00 & 94953.74 & 308998.38 \\
PSOL - CNPJ & 0.139 & 0.278 & 0.111 & 0.028 & 0.222 & 0.083 & 0.083 & 0.028 & 0.028 & 36 & 17.573 & 0.025 &    72.65 & 94953.74 & 288473.38 \\
PSOL - CPF & 0.425 & 0.046 & 0.218 & 0.000 & 0.230 & 0.000 & 0.000 & 0.080 & 0.000 & 87 & 68.605 & 0.000 &    10.00 & 10000.00 & 20525.00 \\
\hline
SD - All & 0.393 & 0.250 & 0.071 & 0.024 & 0.155 & 0.048 & 0.048 & 0.000 & 0.012 & 84 & 24.601 & 0.002 &    44.80 & 3000000.00 & 27990286.64 \\
SD - CNPJ & 0.392 & 0.266 & 0.076 & 0.013 & 0.165 & 0.038 & 0.051 & 0.000 & 0.000 & 79 & 29.100 & 0.000 &  1000.00 & 3000000.00 & 27806667.00 \\
SD - CPF & 1.000 & 0.000 & 0.000 & 0.000 & 0.000 & 0.000 & 0.000 & 0.000 & 0.000 & 1 & 2.322 & 0.970 & 180000.00 & 180000.00 & 180000.00 \\
\hline
\et
\ec
\etab
\end{sidewaystable}

The results in tables \ref{tab:benfordBR} and \ref{tab:benfordCP} show that most declarations have sets of numbers that do not follow Benford's law, while the artificial data generated from the fitted distributions do result in p-values greater than 0. The few exceptions where the declared amounts seem to fit the Benford distribution are those with very small campaigns (few donations), but the artificial data generated in every case seems to render sets of numbers (of the same size as the declared donations) that do fit the Benford distribution, which indicates that the unfitting of data to the Benford prediction is not a bias particular of the donation statistical distributions. The $\chi^2$ values for the most prominent parties are discrepantly high, making it hard to accept that these fluctuations from the expected distribution are merely statistical flaws. Note that the critical value for the $\chi^2$ statistic in order to obtain a p-value of 1\% is around $\chi^2=20.09$ and the p-value drops to zero very fast for higher values of $\chi^2$.

%%%%%%%%%%%%%%%

\section{Conclusions and Overview}

We have analyzed data from Brazil's superior electoral court (TSE) regarding campaign donations and election results. First, the data was fitted to a logistical regression model such that it was possible to significantly determine that the money a candidate receives to run his campaign is a good predictor on whether he is elected or not. Then, assuming that the donations' first digits should naturally follow the Benford distribution, as argued that genuine financial declarations should, we find strong evidence that fraud may have been committed in declarations made by candidates, parties or committees.

Applying well established statistical techniques and results to data concerning Brazil's election campaigns financing and results, it is possible to identify strong evidence that the democratic principles are corrupted: the determining factor on whether a candidate is elected or not is money and there is strong evidence that fraud has been committed in the financial declarations made by the players. If fraud has been committed in these declarations, it is not possible to really determine how the money came to the candidates and therefore it is impossible to know which interests they will be defending once elected.

At this point, we would like to make a small digression. Objective analysis of data (Big Data) has an amazing potential to be beneficial to society, in many different aspects. Close monitorization of individuals medical data could guide public policies that would greatly improve the population's health. Mobile technology allows tracking individuals and monitoring conversations such that crime and terrorism could be more easily solved or, in some cases, even avoided. Consumption data could be used in order to balance the needs and expenses of a population and to optimize the production and industry of a country, minimizing its environmental impacts. Nevertheless, people are usually afraid of sharing data and many strongly argument against it. The reason is simple: no one trust authority. Politicians do not use information for the benefit of the public they rule over; security services usually abuse the power they have and the only thing companies like to optimize is their profits. All the potential science has to be beneficial to society relies on its good public use and therefore, on the leaders the people have. It is of uttermost importance to understand the flaws in our political system such that they can be corrected. In this sense, Brazil has good public information policies and laws. Most government data are made public, and is waiting to be properly analyzed and scrutinized.

%%%%%%%%%%%%%%%%%%%%%%%%%%%

\appendix

\section{Files from TSE}

The data used for the analysis in this paper is publicly available information that can be downloaded from the TSE website. Here we provide information on exactly which files were used to produce the tables and the Md5sums of these files. 

The information concerning the presidential election, comes from files with the \texttt{\_BR} tag (Brazil) in its name. The other tags (\texttt{\_RS}, for example) refer to each one of the 27 Brazilian federal units. For each one of these tags, one can identify the information on donations received (\texttt{receitas\_}) and information about the expanses (\texttt{despesas\_}), since our analysis was about the money the candidates received, we only worked with the former. Moreover, for each category, one identifies three different files, concerning the different players: the parties (\texttt{partidos\_}), committees (\texttt{comites\_}) and candidates (\texttt{candidatos\_}).

For analyzing the presidential campaign, data from the candidates and committees was considered (the two files with the tag \texttt{\_BR}). For the logistic regression model, only data from the candidates was considered. 

In table \ref{tab:md5sum} we show a list of all files used in the analysis and their md5sums.

\btab
\caption{Md5sum for the data files used in the study.}\label{tab:md5sum}
\bc
\bt{c|c}
File & MD5SUM \\
\hline
\texttt{receitas\_candidatos\_2014\_AC.txt}  &   \texttt{eb7981e94258e0ad3a07b6ee6f80bc6b} \\
\texttt{receitas\_candidatos\_2014\_AL.txt}  &   \texttt{12928ae711f9003717ca289bd49f5c7e} \\
\texttt{receitas\_candidatos\_2014\_AM.txt}  &   \texttt{0d49a75f25aebc6a944a5974ef861e8b} \\
\texttt{receitas\_candidatos\_2014\_AP.txt}  &   \texttt{b2d3f5999612306e5bffe38a4f6c237d} \\
\texttt{receitas\_candidatos\_2014\_BA.txt}  &   \texttt{69b592e34ace294c820e13b1d0b2e71d} \\
\texttt{receitas\_candidatos\_2014\_BR.txt}  &   \texttt{271426c7ec86d3dd8814e99408c5db37} \\
\texttt{receitas\_candidatos\_2014\_CE.txt}  &   \texttt{8ddf60c9ae62f13e5efd7b9326a36f1a} \\
\texttt{receitas\_candidatos\_2014\_DF.txt}  &   \texttt{4ba9bdc4f9e897703db801c757b915ff} \\
\texttt{receitas\_candidatos\_2014\_ES.txt}  &   \texttt{10fc502db938054ac6cc3241b57ca582} \\
\texttt{receitas\_candidatos\_2014\_GO.txt}  &   \texttt{a6275ba4cbf919866ee4ee9988db8c7d} \\
\texttt{receitas\_candidatos\_2014\_MA.txt}  &   \texttt{99eeb516159d6095f19ab570061a399a} \\
\texttt{receitas\_candidatos\_2014\_MG.txt}  &   \texttt{0e82ece2e9d260f80cc2bee9d8e5016e} \\
\texttt{receitas\_candidatos\_2014\_MS.txt}  &   \texttt{a839bdeb249e76593ca04b5f13097189} \\
\texttt{receitas\_candidatos\_2014\_MT.txt}  &   \texttt{e18b8f302f611f3b9153c2426690056c} \\
\texttt{receitas\_candidatos\_2014\_PA.txt}  &   \texttt{4a2f22b33f2c66ebf8977f48e4baeba6} \\
\texttt{receitas\_candidatos\_2014\_PB.txt}  &   \texttt{735c1f05dd4091551d2f44dba36c5a04} \\
\texttt{receitas\_candidatos\_2014\_PE.txt}  &   \texttt{18518445913ae55a168765e38fcd9ec2} \\
\texttt{receitas\_candidatos\_2014\_PI.txt}  &   \texttt{3d89349fef1663985de1b2156d0ccb53} \\
\texttt{receitas\_candidatos\_2014\_PR.txt}  &   \texttt{81671fc185ec305af9e699bd2b6e7bfc} \\
\texttt{receitas\_candidatos\_2014\_RJ.txt}  &   \texttt{591f771c495af61be9db0194bb5ff7f4} \\
\texttt{receitas\_candidatos\_2014\_RN.txt}  &   \texttt{9bec87ec41910b0ee0c5b9445ad673d3} \\
\texttt{receitas\_candidatos\_2014\_RO.txt}  &   \texttt{3462f966f3d46492443a6d604c4fea2e} \\
\texttt{receitas\_candidatos\_2014\_RR.txt}  &   \texttt{7f0cc04fafe0ec92a06ba607577f8a0b} \\
\texttt{receitas\_candidatos\_2014\_RS.txt}  &   \texttt{685dd64481b333e783b1c0084e9e8773} \\
\texttt{receitas\_candidatos\_2014\_SC.txt}  &   \texttt{810192175859d3bc8bd05ced22acad8e} \\
\texttt{receitas\_candidatos\_2014\_SE.txt}  &   \texttt{be51cba160864f97f2b1fb216a07e627} \\
\texttt{receitas\_candidatos\_2014\_SP.txt}  &   \texttt{4af106dec3dee88e21bc58f1d20aea6c} \\
\texttt{receitas\_candidatos\_2014\_TO.txt}  &   \texttt{56e3bda5244fe0a3a0dc203feac434d8} \\
\texttt{receitas\_comites\_2014\_BR.txt}  &   \texttt{2920b8be8e66d91e888cc18c3e04a82d} \\
\texttt{receitas\_partidos\_2014\_BR.txt}  &   \texttt{155f374814653e96b9a991cabfc3ca21} 
\et
\ec
\etab

The webpage to obtain these files can be found in

\texttt{http://www.tse.jus.br/eleicoes/estatisticas/repositorio-de-dados-eleitorais},

under the menu ``{\it Prestaçao de Contas}'', by choosing the year 2014. There, a zip file containing all the files in the table (and others) can be downloaded. 

For performing the logistic regression, one also needs to know the situation of each candidate (elected\footnote{There are actually two different ways to be elected: by QP or by ``{\it média}''. We did not differentiate between the two in the analysis.} or not). This information was also obtained from the TSE webpage in CSV file format by filling the form in:

\noindent
{\footnotesize \texttt{http://www.tse.jus.br/eleicoes/estatisticas/estatisticas-candidaturas-2014/estatisticas-eleitorais-2014-resultados}}

and clicking in the link ``{\it Exportar dados}'' (export data).

%%%%%%%%%%%%%%%%%%%%%%%%%%

\section{Python Scripts}

In order to perform the analysis, python classes were programmed in order to read and categorize the information contained in the data files. Also in python, were programmed the routines needed to perform the analysis, fit the distributions and produce plots and tables. All these scripts are available via github in the repository:

\texttt{https://github.com/gamermann/elections}

In order to use the scripts, it is important to have the dependencies properly installed in the system (read more details in the file \texttt{README.txt}) and to have the files from TSE properly downloaded and its location in the computer properly referenced in the scripts. The scripts were programmed to work in a linux system and the tables are compiled in latex, adaptations must be done if the user uses a different system (such as Windows).

%%%%%%%%%%%%%%%

\end{document}